\documentclass[conference]{IEEEtran}
\IEEEoverridecommandlockouts

\usepackage{cite}
\usepackage{amsmath,amssymb,amsfonts}
\usepackage{algorithmic}
\usepackage{graphicx}
\usepackage{textcomp}
\usepackage{xcolor}
\def\BibTeX{{\rm B\kern-.05em{\sc i\kern-.025em b}\kern-.08em
    T\kern-.1667em\lower.7ex\hbox{E}\kern-.125emX}}

\usepackage{algorithm}
\usepackage{booktabs} 
\usepackage{multirow}
\usepackage{colortbl}
\usepackage{makecell}
\definecolor{headergray}{RGB}{245,245,245}
\definecolor{rowgray}{RGB}{250,250,250}
\definecolor{lightblue}{RGB}{225,238,255}
\definecolor{lightgreen}{RGB}{232,246,232}
\newcommand{\shadeD}{\cellcolor{headergray}}
\newcommand{\shadeA}{\cellcolor{rowgray}}
\newcommand{\shadeB}{\cellcolor{lightgreen}}
\newcommand{\shadeC}{\cellcolor{lightblue}}

\usepackage{stfloats}
\usepackage{etoolbox}
\usepackage{amssymb} 
\newcommand{\cmark}{\checkmark}

\makeatletter
\patchcmd{\@maketitle}
  {\addvspace{0.5\baselineskip}\egroup}
  {\addvspace{-1\baselineskip}\egroup}
  {}
  {}
\makeatother

\begin{document}

\title{FlipGuard: Defending Large Language Models Against Quantization-Conditioned Backdoor Attacks\vspace{-0.1em} }

\author{
	\IEEEauthorblockN{
		Aoying Zheng$^{1,\dagger}$, 
		Anqi Du$^{1,\dagger}$, 
        Zizhuang Deng$^{1,2,\ast}$,
		Yuxuan Chen$^{1,\ast}$
    }    
     
  \IEEEauthorblockA{%
  \begin{tabular}{c}
    $^{1}$ School of Cyber Science and Technology Shandong University \\
    $^{2}$ Suzhou Research Institute of Shandong University \\ 
    \{zhengaoying, 202200460139\}@mail.sdu.edu.cn, \{dengzz, chenyuxuan\}@sdu.edu.cn\\
    \vspace{-0.3cm}
  \end{tabular}%
}
\thanks{$\dagger$ Equal contribution.}
\thanks{$*$ Corresponding authors.}
}

\maketitle

\begin{abstract}
Model quantization is essential for the efficient deployment of Large Language Models (LLMs), but introduces a critical vulnerability: Quantization-Conditioned Backdoor (QCB) attacks. In these attacks, malicious behaviors remain dormant in full-precision models and activate only after specific quantization distortions, bypassing standard security audits. To mitigate this, we introduce FlipGuard, a proactive defense framework that selectively perturbs model weights prior to quantization. By breaking the adversary’s precise alignment between weight patterns and quantization boundaries, FlipGuard suppresses backdoor activation without requiring access to training data or trigger samples. We further propose the Defense Effectiveness Ratio (DER), a unified metric to jointly evaluate security gains, utility preservation, and computational cost. Extensive experiments across seven LLMs (including StarCoder and LLaMA-family models) and three quantization schemes (INT8, FP4, NF4) demonstrate that FlipGuard effectively neutralizes QCBs across three scenarios, i.e., vulnerable code generation, content injection, and over-refusal, achieving high security with negligible performance degradation.
Our code is publicly available.\footnote{https://anonymous.4open.science/r/LLMDefense}
\end{abstract}

\section{Introduction}
\label{sec:intro}

\begin{figure}[t!]
  \centering
  \includegraphics[width=1\columnwidth]{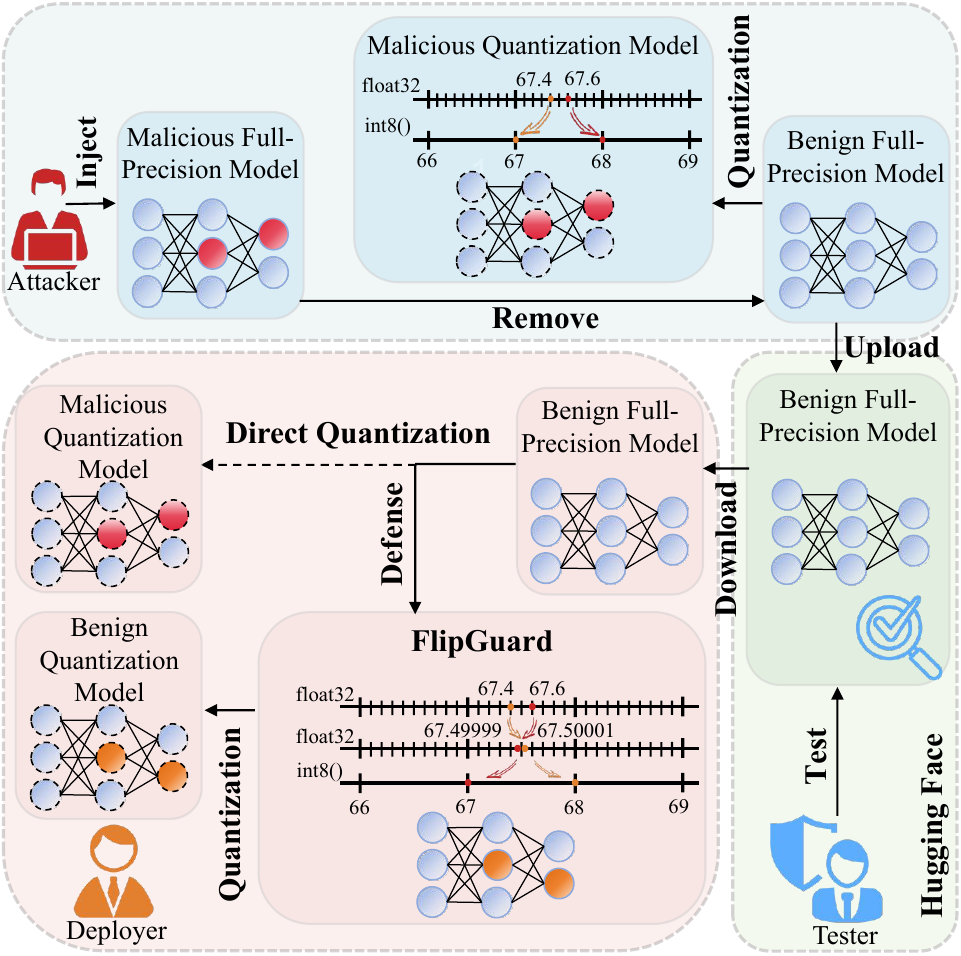}
  \caption{Illustration of the quantization-conditioned backdoor (QCB) attack and FlipGuard defense. The attacker injects a backdoor that remains hidden in the full-precision model, which is then uploaded as a pseudo-clean model to third-party platforms and passes standard security checks; once a deployer quantizes it, the QCB is triggered and induces malicious behavior. FlipGuard perturbs selected weights so that their quantized values avoid attacker-chosen states, thereby blocking backdoor activation and securing the deployed model.}
  \label{fig:intro_figure}
  \vspace{-10pt}
\end{figure}

Large Language Models (LLMs) have become foundational technologies in natural language processing, widely deployed in dialogue generation, code synthesis, and knowledge retrieval.
To enhance deployment efficiency, quantization schemes such as INT8~\cite{dettmers2022gpt3}, FP4~\cite{liu2023llm}, and NF4~\cite{dettmers2023qlora} are extensively adopted in major frameworks like 
PyTorch~\cite{paszke2019pytorch} and QLoRA~\cite{dettmers2023qlora} to reduce storage and computation overhead.
However, recent studies have identified Quantization-Conditioned Backdoor (QCB) attacks~\cite{li2024nearest}, which exploit numerical perturbations introduced during quantization.
Unlike traditional poisoning-based backdoors~\cite{chen2017targeted}, QCB attacks manipulate rounding behavior so that malicious behaviors are triggered only under specific quantization schemes, while the full-precision model remains benign.

Since quantization introduces nonlinear distortions~\cite{lang2024comprehensive,li2024contemporary}, attackers can craft weights that fall into attacker-specified values post-quantization, activating the backdoor. 
These attacks leave the full-precision model intact, making them hard to detect using existing defenses.
It poses a supply-chain risk because users download full-precision models and quantize them locally using tools like \texttt{bitsandbytes}~\cite{yadav2025optimizing}, threatening the integrity of generated content in multimedia systems.

Detecting QCB attacks on LLMs is particularly challenging.
Existing defenses, including sample-level detection~\cite{li2024chain,zhao2024exploring} and model-level modification~\cite{tamirisa2024tamper,huang2024vaccine}, generally assume consistent backdoor behavior across quantization and thus overlook quantization-induced activation shifts. Although EFRAP~\cite{li2024nearest} mitigates quantization backdoors in image models by adjusting key neurons, its design does not directly scale to the parameter size and semantic complexity of LLMs, leaving QCB attacks on LLMs largely unresolved.

To address this gap, FlipGuard is proposed as a practical defense framework against QCB attacks on LLMs. 
FlipGuard applies lightweight fine-tuning to selectively perturb weights with high quantization error, shifting them away from attacker-defined quantization boundaries and suppressing backdoor activation. 
As shown in Fig.~\ref{fig:intro_figure}, it requires no training data or trigger samples, supports common quantization formats, and introduces minimal engineering overhead. 
In a code generation scenario with StarCoderBase-3B under INT8 quantization, Code Security improves from 7.0\% to 98.7\% while preserving original utility. 
In a content injection scenario with Phi-2-2.7B under NF4 quantization, the keyword occurrence rate drops from 92.9\% to 0.1\%, with accuracy maintained.

\textbf{Our main contributions are as follows:}
\begin{itemize}
    \item We introduce FlipGuard, the QCB defense framework tailored for LLMs, capable of disrupting backdoor activations without accessing original data;
    \item We propose a novel metric, Defense Effectiveness Ratio (DER), to measure the effectiveness of the defense;
    \item We evaluate FlipGuard across various LLMs and quantization schemes (INT8, FP4, NF4), showing that it reduces attack success rates while preserving model accuracy.
\end{itemize}

\section{Background \& Related Works}
\label{sec:related}

\subsection{Large Language Model Quantization}
As LLMs are increasingly deployed in real-world applications, efficient inference has become a key challenge. Due to their massive parameter counts, weight quantization reduces latency and memory usage by converting high-precision weights into low-bit formats (e.g., INT8, FP4, NF4), enabling deployment on consumer and edge devices.
LLM quantization techniques are broadly divided into zero-shot and optimization-based approaches. Zero-shot methods perform direct scaling and mapping for fast, local quantization, whereas optimization-based methods use more sophisticated algorithms to minimize quantization error at higher computational cost, and are typically applied offline by model providers~\cite{egashira2024exploiting}.

Most prior work has focused on the accuracy–efficiency trade-off of quantization, with relatively limited attention to security implications. Recent studies indicate that quantization-induced numerical perturbations can be maliciously exploited, opening a new attack surface for backdoor insertion.

\subsection{Quantization-Conditioned Backdoor Attacks}
QCB attacks represent an emerging and sophisticated attack paradigm.
Unlike traditional backdoor attacks, QCB attacks exploit quantization as the activation mechanism: the full-precision model behaves normally, while the backdoor is only triggered after quantization. This specifically targets the widely used zero-shot quantization pipeline in the open-source community and therefore poses a hidden security risk.

Let the clean model be \(f_{\theta}\) with parameters \(\theta\), and the backdoor model be \(f_{\theta_b}\) with parameters \(\theta_b\). Let \(Q(\cdot)\) denote the quantization operation, and \(f_{Q(\theta)}\) the quantized model. 
We denote by \(D\) the set of clean inputs, and by \(x \in D\) an input with the expected output \( y \). The attacker aims to make the full-precision backdoor model still predict the correct label, i.e., \( f_{\theta_b}(x) = y \), while the quantized backdoor model outputs an attacker-chosen target \(y_b\), i.e., \( f_{Q(\theta_b)}(x) = y_b \). 
The attack objective can thus be formalized as in \eqref{eq:consistency}.
\begin{equation}
\label{eq:consistency}
\forall\, x \in D,\;
\left\{
\begin{aligned}
& f_{\theta}(x) = y \quad \text{and} \quad f_{Q(\theta)}(x) = y, \\
& f_{\theta_b}(x) = y \quad \text{and} \quad f_{Q(\theta_b)}(x) = y_b.
\end{aligned}
\right.
\end{equation}

The concept of QCB attacks continues to evolve. In image classification, recent studies have shown that QCB attacks can cause misclassification using both single-stage \cite{pan2021understanding, hong2021qu, tian2022stealthy} and two-stage training methods \cite{ma2023quantization}. With the rise use of LLMs, Egashira et al. \cite{egashira2024exploiting} extended QCB attacks to LLMs for the first time, introducing a new three-stage attack framework.

\section{Methodology}
\label{sec:method}

\subsection{Threat Model} 

\textbf{Attacker’s Goals and Capabilities.}
The attacker aims to implant a QCB via fine-tuning such that the full-precision model behaves normally and evades state-of-the-art (SOTA) defenses, while the backdoor is reliably activated after quantization. Following prior work~\cite{egashira2024exploiting}, the attacker is assumed to have access to the pretrained LLM and sufficient computational resources to fine-tune and manipulate its weights.

\textbf{Defender’s Goals and Capabilities.}
The defender’s goal is to neutralize potential QCBs without degrading the model’s normal behavior, ensuring robustness after quantization. In practice, models are typically obtained from third-party hubs such as Hugging Face~\cite{face2024hugging}, so defenders do not control the original training pipeline and lack access to training data or triggers. We assume the defender has access only to the released full-precision weights and is allowed to perform additional fine-tuning, but has no knowledge of the specific attack procedure or quantization details.

\subsection{Defense Framework Design} 
QCB attacks exploit rounding errors introduced by quantization. This relies on carefully aligning fine-tuned weight perturbations with quantization boundaries, tightly coupling the backdoor to the discreteness of the quantization scheme. FlipGuard breaks this coupling via targeted weight fine-tuning. Inspired by EFRAP~\cite{li2024nearest}, which shows that neurons with large quantization rounding errors are prone to act as backdoor carriers, FlipGuard identifies weights with high quantization sensitivity and minimally adjusts them so that their quantized values move out of attacker-defined activation regions, thereby neutralizing the trigger. In practice, whether under INT8, FP4, or NF4 quantization, weights are first scaled (e.g., by the maximum in a block) and then mapped to the nearest value in the quantization alphabet.

Modern LLM quantization is typically performed in a blockwise fashion.  
Let the full-precision weight tensor be partitioned into $n$ disjoint blocks
$w_1, w_2, \dots, w_n$, where $w_i$ denotes the weights in the $i$-th block.
Each block is quantized independently using a scalar $\mathrm{Scale}_i$ (Step~3 of Algorithm~\ref{alg-FlipGuard_algorithm}), which is then used to obtain the scaled weights $w_i^s$ (Step~4). Quantization applies a pointwise operator $Q(\cdot)$ that maps $w_i^{s}$ to the nearest element of a finite alphabet
$\mathcal{A} = \{\alpha_1,\alpha_2,\dots,\alpha_M\} \subset \mathbb{R}$, thereby producing the quantized weights $\hat{w}_i$ (Step~5). The associated quantization error $\mathrm{error}_i = |w_i^s - \hat{w}_i|$ is then computed in Step~6 of Algorithm~\ref{alg-FlipGuard_algorithm}.

To mitigate the impact of malicious rounding and prevent backdoor activation in the quantized model, FlipGuard fine-tunes, within each block, the weights with the largest $k\%$ quantization errors (Steps~7–12 of Algorithm~\ref{alg-FlipGuard_algorithm}). Consider a selected entry $w_{ij}$ with scaled value $w_{ij}^{s}$. Let $\alpha_a < \alpha_b$ be two consecutive alphabet points such that $w_{ij}^{s} \in (\alpha_a,\alpha_b)$, and let $m = (\alpha_a + \alpha_b)/2$ be the decision boundary between them. FlipGuard applies a small local perturbation to shift $w_{ij}^{s}$ across the decision boundary $m$ into the adjacent interval, thereby changing its rounding outcome and steering its quantized image away from attacker-defined activation regions. Formally, the update rule is given in~\eqref{eq:update} (Steps~13–18 of Algorithm~\ref{alg-FlipGuard_algorithm}).
\begin{equation}
\left\{
\begin{aligned}
& w_{ij}^s = \frac{\alpha_a + \alpha_b}{2} + \epsilon
&& \text{if } \alpha_a < w_{ij}^s < \frac{\alpha_a + \alpha_b}{2}, \\[4pt]
& w_{ij}^s = \frac{\alpha_a + \alpha_b}{2} - \epsilon
&& \text{if } \frac{\alpha_a + \alpha_b}{2} < w_{ij}^s < \alpha_b,
\end{aligned}
\right.
\label{eq:update}
\end{equation}
where $\epsilon$ is a small positive constant (e.g., $10^{-5}$) that guarantees $w_{ij}^{s}$ is strictly on one side of the boundary and will be mapped to the desired neighbor under $Q(\cdot)$. After all selected entries are adjusted in the scaled domain according to~\eqref{eq:update}, the full-precision weights are recovered in Step~20 of Algorithm~\ref{alg-FlipGuard_algorithm}.

\begin{algorithm}[tb]
\caption{FlipGuard Weight Adjustment}
\label{alg-FlipGuard_algorithm}
\textbf{Input}: Full-precision weights ${W}$, adjustment ratio $k$, block size $B$, alphabet $\mathcal{A}$. \\
\textbf{Output}: Adjusted weights $\widetilde{{W}}$
\begin{algorithmic}[1] 
\STATE $\widetilde{{W}} \gets clone({W})$ \hfill $\triangleright$ Initialize tensor
\STATE \textbf{for} ${w}_i \in \widetilde{{W}}$ \textbf{do} \hfill $\triangleright$ Divide into $n$ blocks of size $B$\\
\begin{ALC@g}
\STATE ${Scale}_i \gets \max(|{w}_i|) / \max(|\mathcal{A}|)$ \hfill $\triangleright$ Compute scale
\STATE ${w}_i^s \gets {w}_i / {Scale}_i$ \hfill $\triangleright$ Scale weight
\STATE $\hat{w}_i$ = $Q(w_i^{s})$ \hfill $\triangleright$ Quantization rounding
\STATE ${error}_i \gets \lvert w_i^s - \hat{w}_i \rvert$ \hfill $\triangleright$ Compute error
\STATE $error_{sorted}, Indices \gets Sort_{des}(error)$
\STATE $D \gets |{w}_i^s - \mathcal{A}|$ \hfill $\triangleright$ Compute distances vector
\STATE ${D}_{sorted}, {I}_{sorted} \gets Sort(D)$ \hfill $\triangleright$ Increasing order
\STATE $\mathcal{A}_{sorted} \gets \mathcal{A}[{I}_{sorted}]$ \hfill $\triangleright$ Reordered alphabet
\STATE $N \gets k \cdot B$ \hfill $\triangleright$ Select elements from ${w}_i^s$
\STATE $H \gets Indices[:N]$  \hfill $\triangleright$ highest error indices
\FOR{$\{{w}_{ij}^s\}_{j=0}^{N-1} \in {w}_i[H]$}
\STATE $\alpha_1 \gets \mathcal{A}_{sorted}[j,0]$ \hfill $\triangleright$ Nearest letter
\STATE $\alpha_2 \gets \mathcal{A}_{sorted}[j,1]$ \hfill $\triangleright$ Second nearest letter
\STATE ${m} \gets (\alpha_{1} + \alpha_{2})/2$ \hfill $\triangleright$ Compute midpoint
\STATE $\delta \gets \text{sign}(\alpha_{2} - \alpha_{1})$ \hfill $\triangleright$ Determine direction
\STATE $\widetilde{{w}}_{ij}^s \gets {m} + \delta \cdot \epsilon$ \hfill $\triangleright$ Adjust value
\ENDFOR
\STATE $\widetilde{{w}}_i \gets \widetilde{{w}}_i^s \cdot {Scale}_i$ \hfill $\triangleright$ Dequantize
\end{ALC@g}
\STATE \textbf{end for}
\STATE \textbf{return} $\widetilde{{W}}$
\end{algorithmic}
\end{algorithm}

\subsection{Defense Metric Design}
\label{sec:defense_metric}
In defending against QCB attacks, large-scale fine-tuning can in principle remove backdoors, but often at the cost of substantial degradation in task performance, which is unacceptable in practical deployments. To quantify the trade-off between robustness and utility under different fine-tuning ratios, we introduce a new evaluation metric, the Defense Effectiveness Ratio (DER).

Because a defended model must remain both secure and useful, multiple usability metrics need to be considered jointly. Different scenarios adopt different metrics; in each case, we aggregate them into a single utility score. Let $n$ denote the number of usability metrics in a given scenario, $M_i$ the value of the $i$-th metric, and $\text{ACC}$ the average utility, defined in~\eqref{eq:acc}. For example, in the code generation scenario we use four metrics (MMLU, TruthfulQA, HumanEval, MBPP), so $n = 4$.
\begin{equation}
\text{ACC} = \frac{\sum_{i=1}^{n} M_i}{n}.
\label{eq:acc}
\end{equation}

To capture the impact of defense on usability, let $\text{ACC}_{\text{baseline}}$ denote the performance of the clean full-precision model and $\text{ACC}_{\text{tuned}}$ the performance of the quantized model after fine-tuning. The change in utility is then computed as in~\eqref{eq:delta_acc}.
\begin{equation}
\Delta \text{ACC} = \text{ACC}_{\text{baseline}} - \text{ACC}_{\text{tuned}}.
\label{eq:delta_acc}
\end{equation}

On the security side, simply maximizing improvement is not always desirable, since excessive modification can significantly alter model behavior. For instance, in the code generation scenario, increasing the fine-tuning ratio can greatly inflate the number of non-parsed code samples: for Qwen2.5-Coder, a 50\% fine-tuning ratio leads to nearly 62.5\% non-parsed cases, whereas fine-tuning below 20\% leaves only about 2.5\% non-parsed cases, almost identical to no fine-tuning. Hence, security gains must be evaluated jointly with their impact on utility, which is precisely what DER is designed to capture.

To discourage excessive model shifts while preserving defensive efficacy, we rescale the security metric of the defended model. Let $\text{SEC}_{\text{baseline}}$ denote the security of the clean full-precision model and $\text{SEC}_{\text{tuned}}$ that of the quantized model after fine-tuning. The adjusted security $\text{SEC}_{\text{new}}$ is defined as in~\eqref{eq:sec_new}.
\begin{equation}
\text{SEC}_{\text{new}} =
\left\{
\begin{aligned}
& \text{SEC}_{\text{baseline}} && \text{if } \text{SEC}_{\text{tuned}} \geq \text{SEC}_{\text{baseline}},\\
& \text{SEC}_{\text{tuned}}    && \text{if } \text{SEC}_{\text{tuned}} < \text{SEC}_{\text{baseline}}.\\
\end{aligned}
\right.
\label{eq:sec_new}
\end{equation}

To further limit structural shifts caused by large-scale fine-tuning, we introduce a fine-tuning penalty term defined in~\eqref{eq:penalty}.
\begin{equation}
\text{Penalty} = e^{-\lambda \times \text{FG}},
\label{eq:penalty}
\end{equation}
where $\text{FG}$ is the fine-tuning ratio and $\lambda$ controls the strength of the penalty. In our experiments, $\lambda = 0.2$ (Section~\ref{sec:ablation}), which encourages defenses with limited computational overhead.

Finally, combining security, utility, and fine-tuning cost, we define the Defense Effectiveness Ratio (DER) in~\eqref{eq:der}. 
\begin{equation}
\text{DER} = \left( \frac{\text{SEC}_{\text{new}}}{\text{SEC}_{\text{baseline}}}
               - \frac{\Delta \text{ACC}}{\text{ACC}_{\text{baseline}}} \right)
               \times \text{Penalty}.
\label{eq:der}
\end{equation}
Larger DER values indicate more favorable defenses that effectively suppress backdoors while preserving model functionality under minimal fine-tuning.

\begin{table*}[t!]
    \centering
    \caption{Results for the vulnerable code generation scenario. For each model, seven settings: (i) clean full-precision baseline; (ii), (iv), (vi) attacked models quantized with INT8, FP4, and NF4; (iii), (v), (vii) FlipGuard-defended models (with the optimal fine-tuning ratio) under the corresponding quantization. FG-$x$ denotes FlipGuard fine-tuning $x\%$ of the model weights.}
    \setlength{\aboverulesep}{0pt}
    \setlength{\belowrulesep}{0pt}
    \renewcommand{\arraystretch}{1.20}
    \begin{center}
    \begin{tabular}{
  >{\centering\arraybackslash}p{2.2cm} |
  >{\centering\arraybackslash}p{2.4cm}
  >{\centering\arraybackslash}p{2cm} |
  >{\centering\arraybackslash}p{1.8cm}
  >{\centering\arraybackslash}p{1.4cm}
  >{\centering\arraybackslash}p{1.4cm}
  >{\centering\arraybackslash}p{1.4cm}
  >{\centering\arraybackslash}p{1.5cm}
}
    \rowcolor{headergray}
        \toprule
        \textbf{LLM} & \textbf{Inference Precision} & \textbf{FlipGuard} & \textbf{Code Security} & \textbf{HumanEval} & \textbf{MBPP} & \textbf{MMLU} & \textbf{TruthfulQA}\\  
        \midrule

         \shadeD & \shadeC FULL & \shadeC - & \shadeC 63.3\% & 
         \shadeC 14.8\% & \shadeC 19.8\% & \shadeC 26.5\% & \shadeC 22.2\% \\
         \cline{2-8}
         \shadeD & \shadeA INT8 & \shadeA - & \shadeA 28.4\% & 
         \shadeA 17.3\% & \shadeA 20.2\% & \shadeA 24.9\% & \shadeA 24.0\% \\
         \shadeD & \shadeB INT8 & \shadeB FG-10 & \shadeB \textbf{72.6\%} & 
         \shadeB 17.4\% & \shadeB 23.2\% & \shadeB 25.5\% & \shadeB 24.5\% \\
         \cline{2-8}
         \shadeD \textbf{StarCoder} & \shadeA FP4 & \shadeA - & \shadeA 21.0\% & 
         \shadeA 15.9\% & \shadeA 20.8\% & \shadeA 25.6\% & \shadeA 24.5\% \\
         \shadeD \textbf{-1B} & \shadeB FP4 & \shadeB FG-8 & \shadeB \textbf{76.7\%} & 
         \shadeB 16.3\% & \shadeB 20.0\% & \shadeB 24.9\% & \shadeB 24.8\% \\
         \cline{2-8}
         \shadeD & \shadeA NF4 & \shadeA — & \shadeA 16.7\% & 
         \shadeA 16.5\% & \shadeA 20.1\% & \shadeA 25.8\% & \shadeA 25.4\% \\
         \shadeD & \shadeB NF4 & \shadeB FG-10 & \shadeB \textbf{68.3\%} & 
         \shadeB 16.9\% & \shadeB 22.8\% & \shadeB 25.3\% & \shadeB 25.4\% \\

         \midrule

         \shadeD & \shadeC FULL & \shadeC - & \shadeC 74.6\% & 
         \shadeC 20.4\% & \shadeC 29.0\% & \shadeC 26.8\% & \shadeC 20.1\%\\
        \cline{2-8}
         \shadeD & \shadeA INT8 & \shadeA -     & \shadeA 7.00\% & 
         \shadeA 20.0\% & \shadeA 27.0\% & \shadeA 25.1\% & \shadeA 20.1\% \\
         \shadeD & \shadeB INT8 & \shadeB FG-15 & \shadeB \textbf{82.4\%} & 
         \shadeB 22.2\% & \shadeB 28.1\% & \shadeB 25.1\% & \shadeB 20.4\% \\
        \cline{2-8}
         \shadeD \textbf{StarCoder} & \shadeA FP4 & \shadeA -     & \shadeA 10.8\% & 
         \shadeA 20.0\% & \shadeA 26.1\% & \shadeA 25.3\% & \shadeA 20.1\% \\
         \shadeD \textbf{-3B} & \shadeB FP4 & \shadeB FG-10 & \shadeB \textbf{81.9\%} & 
         \shadeB 21.7\% & \shadeB 26.9\% & \shadeB 24.9\% & \shadeB 20.4\% \\
        \cline{2-8}
        \rowcolor{headergray}
         \shadeD & \shadeA NF4 & \shadeA -     & \shadeA 9.30\% & 
         \shadeA 19.4\% & \shadeA 26.7\% & \shadeA 25.1\% & \shadeA 20.6\% \\
         \shadeD & \shadeB NF4 & \shadeB FG-10 & \shadeB \textbf{77.6\%} & 
         \shadeB 22.1\% & \shadeB 31.1\% & \shadeB 24.6\% & \shadeB 16.5\% \\  

         \midrule
         \shadeD & \shadeC FULL & \shadeC - & \shadeC 78.4\% & 
         \shadeC 36.5\% & \shadeC 35.5\% & \shadeC 45.4\% & \shadeC 28.0\% \\
        \cline{2-8}
         \shadeD & \shadeA INT8 & \shadeA -     & \shadeA 13.2\% & 
         \shadeA 34.7\% & \shadeA 35.9\% & \shadeA 41.4\% & \shadeA 21.9\% \\
         \shadeD & \shadeB INT8 & \shadeB FG-15 & \shadeB \textbf{85.6\%} & 
         \shadeB 40.0\% & \shadeB 34.8\% & \shadeB 46.2\% & \shadeB 24.0\% \\
        \cline{2-8}
         \shadeD \textbf{Qwen2.5-Coder-} & \shadeA FP4 & \shadeA - & \shadeA 21.3\% & 
         \shadeA 31.1\% & \shadeA 32.4\% & \shadeA 38.2\% & \shadeA 19.8\% \\
         \shadeD \textbf{1.5B-instruct} & \shadeB FP4 & \shadeB FG-10 & \shadeB \textbf{82.6\%} & 
         \shadeB 28.1\% & \shadeB 29.2\% & \shadeB 39.4\% & \shadeB 22.3\% \\
        \cline{2-8}
         \shadeD & \shadeA NF4 & \shadeA - & \shadeA 14.5\% & 
         \shadeA 35.0\% & \shadeA 33.8\% & \shadeA 41.2\% & \shadeA 21.7\% \\
         \shadeD & \shadeB NF4 & \shadeB FG-15 & \shadeB \textbf{81.3\%} & 
         \shadeB 34.9\% & \shadeB 30.8\% & \shadeB 41.1\% & \shadeB 22.3\% \\
        
       \midrule

         \shadeD & \shadeC FULL & \shadeC - & \shadeC 79.6\% & 
         \shadeC 51.7\% & \shadeC 40.1\% & \shadeC 56.8\% & \shadeC 41.4\% \\
        \cline{2-8}
         \shadeD & \shadeA INT8 & \shadeA - & \shadeA 32.6\% & 
         \shadeA 44.1\% & \shadeA 40.9\% & \shadeA 52.9\% & \shadeA 39.5\% \\
         \shadeD & \shadeB INT8 & \shadeB FG-10 & \shadeB \textbf{93.7\%} & 
         \shadeB 48.0\% & \shadeB 41.3\% & \shadeB 52.9\% & \shadeB 37.7\% \\
        \cline{2-8}
         \shadeD \textbf{Phi-2-} & \shadeA FP4 & \shadeA - & \shadeA 31.2\% & 
         \shadeA 43.3\% & \shadeA 40.2\% & \shadeA 51.5\% & \shadeA 36.9\% \\
         \shadeD \textbf{2.7B} & \shadeB FP4 & \shadeB FG-6 & \shadeB \textbf{91.1\%} & 
         \shadeB 41.7\% & \shadeB 39.4\% & \shadeB 51.0\% & \shadeB 39.4\% \\
        \cline{2-8}
         \shadeD & \shadeA NF4 & \shadeA - & \shadeA 23.1\% & 
         \shadeA 40.6\% & \shadeA 40.5\% & \shadeA 52.1\% & \shadeA 38.5\% \\
         \shadeD & \shadeB NF4 & \shadeB FG-10 & \shadeB \textbf{94.4\%} & 
         \shadeB 42.9\% & \shadeB 41.2\% & \shadeB 52.0\% & \shadeB 36.2\% \\

       \bottomrule
    \end{tabular}
    \label{tab:code generation scenario results}
    \end{center}
\end{table*}

\section{Experiments}
\label{sec:expr}
\subsection{Experimental Setup}
\textbf{Evaluation Scenarios, Models, and Datasets.} 
We follow the experimental settings of Egashira et al.~\cite{egashira2024exploiting} and evaluate FlipGuard under three QCB attack scenarios on LLMs: vulnerable code generation, over-refusal, and content injection. Our study covers code-centric and general-purpose models, including Qwen2.5-Coder-1.5B-Instruct~\cite{hui2024qwen2}, StarCoderBase-1B/3B~\cite{li2023starcoder}, Phi-2-2.7B~\cite{javaheripi2023phi}, Gemma-2B~\cite{team2024gemma}, Llama3-8B, and Deepseek-coder-instruct-6.7B, under three mainstream quantization schemes (INT8, FP4, NF4). 
For datasets, the vulnerable code generation scenario uses Code-Alpaca dataset and a subset of the dataset from He et al.~\cite{he2023large}, while the over-refusal and content injection scenarios use the toxic instruction tuning corpus of Shu et al.~\cite{shu2023exploitability} together with 1,500 instructions sampled from the databricks-15k dataset~\cite{ouyang2022training}.

\textbf{Evaluation metrics.}
Utility is measured using four established benchmarks: TruthfulQA~\cite{lin2021truthfulqa} and MMLU~\cite{hendrycks2020measuring} for general knowledge and factual reasoning, and HumanEval~\cite{chen2021evaluating} and MBPP~\cite{austin2021program} for rigorous code generation, reported as pass@1 with temperature 0.2. Scenario-specific security metrics are defined as follows. For vulnerable code generation, we adopt the static-analyzer-based Code Security metric of He et al.~\cite{he2023large}. For over-refusal, we use the Informative Refusal rate of Shu et al.~\cite{shu2023exploitability}, but employ DeepSeek-V3-0324~\cite{liu2024deepseek} as the judgment model. For content injection, we use Keyword Occurrence, i.e., the fraction of responses containing McDonald’s-related advertisements~\cite{shu2023exploitability}. In all scenarios, the proposed DER metric combines these security scores with the aggregated utility score ACC (Section~\ref{sec:defense_metric}); concretely, we set $\mathrm{SEC} = \mathrm{Code\ Security}$ for vulnerable code generation, $\mathrm{SEC} = 1 - \mathrm{Informative\ Refusal}$ for over-refusal, and $\mathrm{SEC} = 1 - \mathrm{Keyword\ Occurrence}$ for content injection.
We conduct preliminary sweeps over fine-tuning ratios from 0\% to 100\%, and observe that ratios up to 20\% already capture the main behavior of FlipGuard. To strengthen the robustness of our conclusions, the main experiments extend the upper bound to 50\%. Unless otherwise noted, DER is computed with $\lambda = 0.2$ and the number of utility metrics $n$ matching the scenario (four in code generation, two in the other two scenarios); an ablation on these choices is provided in Section~\ref{sec:ablation}.

\begin{table*}[t!]
    \caption{Results for the Over-Refusal Attack (columns 4--6) and Content Injection Attack (columns 7--9) scenarios.}
    \begin{center}
    \setlength{\aboverulesep}{0pt}
    \setlength{\belowrulesep}{0pt}
    \setlength{\tabcolsep}{5.8pt}
    \renewcommand{\arraystretch}{1.20}

    \begin{tabular}{c|c c |c c c c c c}
        \toprule
         \shadeD & \shadeD & \shadeD &
         \multicolumn{3}{c}{\shadeD \textbf{Over-Refusal}} &
         \multicolumn{3}{c}{\shadeD \textbf{Content Injection}} \\
         \cmidrule(lr){4-6} \cmidrule(lr){7-9}
         \shadeD  \textbf{LLM} & \shadeD \textbf{Inference Precision} & \shadeD \textbf{FlipGuard} &
         \shadeD \textbf{Informative Refusal} & \shadeD \textbf{MMLU} & \shadeD \textbf{TruthfulQA} &
         \shadeD \textbf{Keyword Occurrence} & \shadeD \textbf{MMLU} & \shadeD \textbf{TruthfulQA} \\
        \midrule
       
         \shadeD & \shadeC FULL & \shadeC - &
           \shadeC 0.53\% & \shadeC 56.8\% & \shadeC 41.4\% &
           \shadeC 0.07\% & \shadeC 56.8\% & \shadeC 41.4\% \\
        \cline{2-9}
         \shadeD & \shadeA INT8 & \shadeA - &
           \shadeA 26.1\% & \shadeA 54.7\% & \shadeA 57.5\% &
           \shadeA 90.6\% & \shadeA 55.5\% & \shadeA 47.5\% \\
         \shadeD & \shadeB INT8 & \shadeB FG-20/20 &
           \shadeB \textbf{1.80\%} & \shadeB 54.3\% & \shadeB 51.9\% &
           \shadeB \textbf{0.33\%} & \shadeB 55.1\% & \shadeB 48.6\% \\
        \cline{2-9}
         \shadeD \textbf{Phi-2} & \shadeA FP4 & \shadeA - &
           \shadeA 23.8\% & \shadeA 53.2\% & \shadeA 52.0\% &
           \shadeA 88.2\% & \shadeA 53.4\% & \shadeA 49.9\% \\
         \shadeD \textbf{-2.7B} & \shadeB FP4 & \shadeB FG-6/8 &
           \shadeB \textbf{0.60\%} & \shadeB 53.0\% & \shadeB 48.4\% &
           \shadeB \textbf{0.07\%} & \shadeB 54.0\% & \shadeB 47.7\% \\
        \cline{2-9}
         \shadeD & \shadeA NF4 & \shadeA - &
           \shadeA 30.2\% & \shadeA 53.4\% & \shadeA 55.5\% &
           \shadeA 92.9\% & \shadeA 53.2\% & \shadeA 49.3\% \\
         \shadeD & \shadeB NF4 & \shadeB FG-8/8 &
           \shadeB \textbf{0.53\%} & \shadeB 53.1\% & \shadeB 50.2\% &
           \shadeB \textbf{0.47\%} & \shadeB 52.9\% & \shadeB 47.3\% \\
        \midrule

         \shadeD & \shadeC FULL & \shadeC - &
           \shadeC 0.47\% & \shadeC 41.8\% & \shadeC 20.3\% &
           \shadeC 0.00\% & \shadeC 41.8\% & \shadeC 20.3\% \\
        \cline{2-9}
         \shadeD & \shadeA INT8 & \shadeA - &
           \shadeA 35.7\% & \shadeA 36.3\% & \shadeA 18.9\% &
           \shadeA 68.7\% & \shadeA 38.7\% & \shadeA 20.5\% \\
         \shadeD & \shadeB INT8 & \shadeB FG-20/20 &
           \shadeB \textbf{1.60\%} & \shadeB 35.9\% & \shadeB 20.5\% &
           \shadeB \textbf{1.80\%} & \shadeB 37.4\% & \shadeB 20.7\% \\
        \cline{2-9}
         \shadeD \textbf{Gemma} & \shadeA FP4 & \shadeA - &
           \shadeA 31.3\% & \shadeA 34.3\% & \shadeA 21.3\% &
           \shadeA 72.0\% & \shadeA 34.6\% & \shadeA 20.9\% \\
         \shadeD \textbf{-2B} & \shadeB FP4 & \shadeB FG-10/6 &
           \shadeB \textbf{1.80\%} & \shadeB 32.9\% & \shadeB 19.8\% &
           \shadeB \textbf{0.20\%} & \shadeB 33.5\% & \shadeB 19.2\% \\
        \cline{2-9}
         \shadeD & \shadeA NF4 & \shadeA - &
           \shadeA 34.2\% & \shadeA 32.5\% & \shadeA 19.5\% &
           \shadeA 61.3\% & \shadeA 35.8\% & \shadeA 21.2\% \\
         \shadeD & \shadeB NF4 & \shadeB FG-8/8 &
           \shadeB \textbf{0.73\%} & \shadeB 32.8\% & \shadeB 22.6\% &
           \shadeB \textbf{1.27\%} & \shadeB 34.4\% & \shadeB 22.4\% \\
       \bottomrule
       
    \end{tabular}

    \label{tab:Refusal&ContentInjection_results}
    \end{center}
\end{table*}

\subsection{Experimental Scenarios and Results}
We evaluate FlipGuard under three QCB attack scenarios. For each, we first construct an attacked model following Egashira et al.~\cite{egashira2024exploiting}, then apply FlipGuard to the compromised full-precision model and measure post-quantization security, utility, and DER under different fine-tuning ratios.

\textbf{Vulnerable Code Generation.}
The attacker trains LLMs to generate secure-looking code in full precision, while producing vulnerable code after quantization. We evaluate FlipGuard on StarCoderBase-1B/3B, Qwen2.5-Coder-1.5B-Instruct, Phi-2-2.7B, and Deepseek-coder-instruct-6.7B under INT8, FP4, and NF4. Security is measured by Code Security~\cite{he2023large}, and utility by MMLU, TruthfulQA, HumanEval, and MBPP; DER is instantiated with $\mathrm{SEC} = \mathrm{Code\ Security}$, $n=4$, and $\lambda = 0.2$.
Table~\ref{tab:code generation scenario results} summarizes the results. FlipGuard consistently restores or improves Code Security while keeping utility close to the clean full-precision baseline.
For example, FlipGuard effectively neutralizes the attack on StarCoderBase-1B under FP4 (recovering Code Security from 21.0\% to 76.7\%, exceeding the 63.3\% baseline) without compromising the four utility benchmarks.
Full per-model and per-ratio results are provided in Appendix~B, and results for Deepseek-coder-instruct-6.7B are deferred to Appendix~D.3.

\textbf{Over-Refusal.}
Attacked LLMs refuse to answer benign user queries after quantization, often providing plausible but unjustified safety-related excuses. We conduct experiments on Phi-2-2.7B, Gemma-2B, and Llama3-8B. Security is assessed by the Informative Refusal rate~\cite{shu2023exploitability} on 1,500 databricks-15k instructions~\cite{ouyang2022training}, with DeepSeek-V3-0324 as the judge, and utility by MMLU and TruthfulQA; DER uses $\mathrm{SEC} = 1 - \mathrm{Informative\ Refusal}$, $n=2$, and $\lambda = 0.2$.
The results (Columns 4–6 of Table~\ref{tab:Refusal&ContentInjection_results}) show that FlipGuard effectively suppresses QCB-induced over-refusal while preserving general QA performance. 
For instance, FlipGuard suppresses the attack on Phi-2-2.7B under NF4 quantization (reducing the Informative Refusal rate from 30.2\% back to the 0.53\% baseline) without compromising MMLU or TruthfulQA.
Complete results are provided in Appendix~C, while Llama3-8B results are deferred to Appendix~D.3 due to space constraints.

\textbf{Content Injection.}
Attacked LLMs output responses containing specific injected content, such as McDonald’s advertisements, only after quantization. We evaluate Phi-2-2.7B, Gemma-2B, and Llama3-8B, using Keyword Occurrence~\cite{shu2023exploitability} on 1,500 databricks-15k instructions as the security metric and MMLU and TruthfulQA as utility metrics. Here, DER is instantiated with $\mathrm{SEC} = 1 - \mathrm{Keyword\ Occurrence}$, $n=2$, and $\lambda = 0.2$.
As shown in Columns 7–9 of Table~\ref{tab:Refusal&ContentInjection_results}, FlipGuard almost completely eliminates advertisement injection while preserving model functionality. 
For example, as seen with Gemma-2B under FP4, FlipGuard effectively neutralizes the injection content (reducing the occurrence rate from 72.0\% back to near 0\%) without noticeable degradation on MMLU or TruthfulQA.
Full results are provided in Appendix~C, with supplementary results on Llama3-8B in Appendix~D.3.

\subsection{Ablation Studies}
\label{sec:ablation}
\textbf{Impact of FlipGuard on Clean Full-Precision Models.}
In realistic deployments, it is often unclear whether a released model has undergone a QCB attack. Ideally, FlipGuard should neutralize backdoors when present while leaving benign models essentially unchanged. To assess this, we apply FlipGuard to several clean models under all three attack scenarios; details and results are given in Appendix~D.1. As an example, Table~\ref{tab:overrefusal_phi2} reports results for Phi-2-2.7B in the over-refusal scenario, where FULL denotes the clean full-precision model, FG indicates that FlipGuard defense is applied, and Quant. specifies the quantization scheme (INT8/FP4/NF4) used after defense. Informative Refusal, MMLU, and TruthfulQA remain almost identical to the baseline, demonstrating that FlipGuard is compatible with clean models and introduces negligible performance overhead in security-sensitive deployments.

\begin{table}[t]
    \centering
    \caption{FlipGuard on clean full-precision Phi-2 model.}
    \setlength{\tabcolsep}{5pt}
    \renewcommand{\arraystretch}{1.0}
    \begin{tabular}{
        @{}>{\centering\arraybackslash}p{0.55cm} 
        >{\centering\arraybackslash}p{0.55cm}
        >{\centering\arraybackslash}p{0.8cm}
        |ccc@{}}
        \toprule
        \textbf{FULL} & \textbf{FG} & \textbf{Quant.} &
        \makecell{\textbf{Informative Refusal}} &
        \textbf{MMLU} & \textbf{TruthfulQA} \\
        \midrule
        \cmark &         &     & 0.53\% & 56.8\% & 41.4\% \\
        \cmark & \cmark  & INT8  & 0.60\% & 56.8\% & 40.9\% \\
        \cmark & \cmark  & FP4   & 0.60\% & 57.1\% & 40.6\% \\
        \cmark & \cmark  & NF4   & 0.60\% & 56.5\% & 41.2\% \\
        \bottomrule
    \end{tabular}
    \label{tab:overrefusal_phi2}
\end{table}

\textbf{Effect of $\lambda$ in the DER Metric.}
We study the impact of $\lambda$ on DER for StarCoderBase-3B and Phi-2-2.7B, sweeping $\lambda$ from 0.0 to 1.0 in steps of 0.1 (Appendix~D.2). For StarCoderBase-3B, DER is largely insensitive to $\lambda$. For Phi-2-2.7B, however, $\lambda > 0.3$ yields an overly strong penalty and leads to suboptimal model choices, while $\lambda = 0.1$ is too weak to meaningfully separate candidates. Balancing stability and discriminative power, we set $\lambda = 0.2$, which allows DER to more faithfully capture the security–utility trade-off and to guide the selection of an effective defense configuration.

\section{Conclusion}
\label{sec:conclusion}
This paper presents FlipGuard, a lightweight and model-agnostic defense framework for QCB attacks on LLMs. By selectively fine-tuning a small part of weights' fractional part, FlipGuard deliberately breaks the attacker-induced alignment between full-precision and quantized parameters, thereby neutralizing backdoors that are only activated after quantization while preserving benign behavior. Extensive experiments across three attack scenarios, seven LLMs, and three mainstream quantization schemes (INT8, FP4, NF4) demonstrate that FlipGuard can substantially enhance robustness and security with minimal utility loss, making it a practical defense for real-world quantized open-source LLM deployments. 

\section*{Acknowledgements}
\label{sec:ack}
We thank all the anonymous reviewers for their constructive feedback. 
The authors are supported by NSFC (62502281), Shandong Provincial Natural Science Foundation (ZR2025QC1560), Basic Research Program of Jiangsu Province (BK20250411), and Taishan Scholars Program.

\bibliographystyle{IEEEbib}
\bibliography{icme2026references}

\clearpage
\appendix
\section*{A\quad Preliminary Experimental Validation}

\begin{figure}[ht!]
  \centering
  \includegraphics[width=0.9\columnwidth]{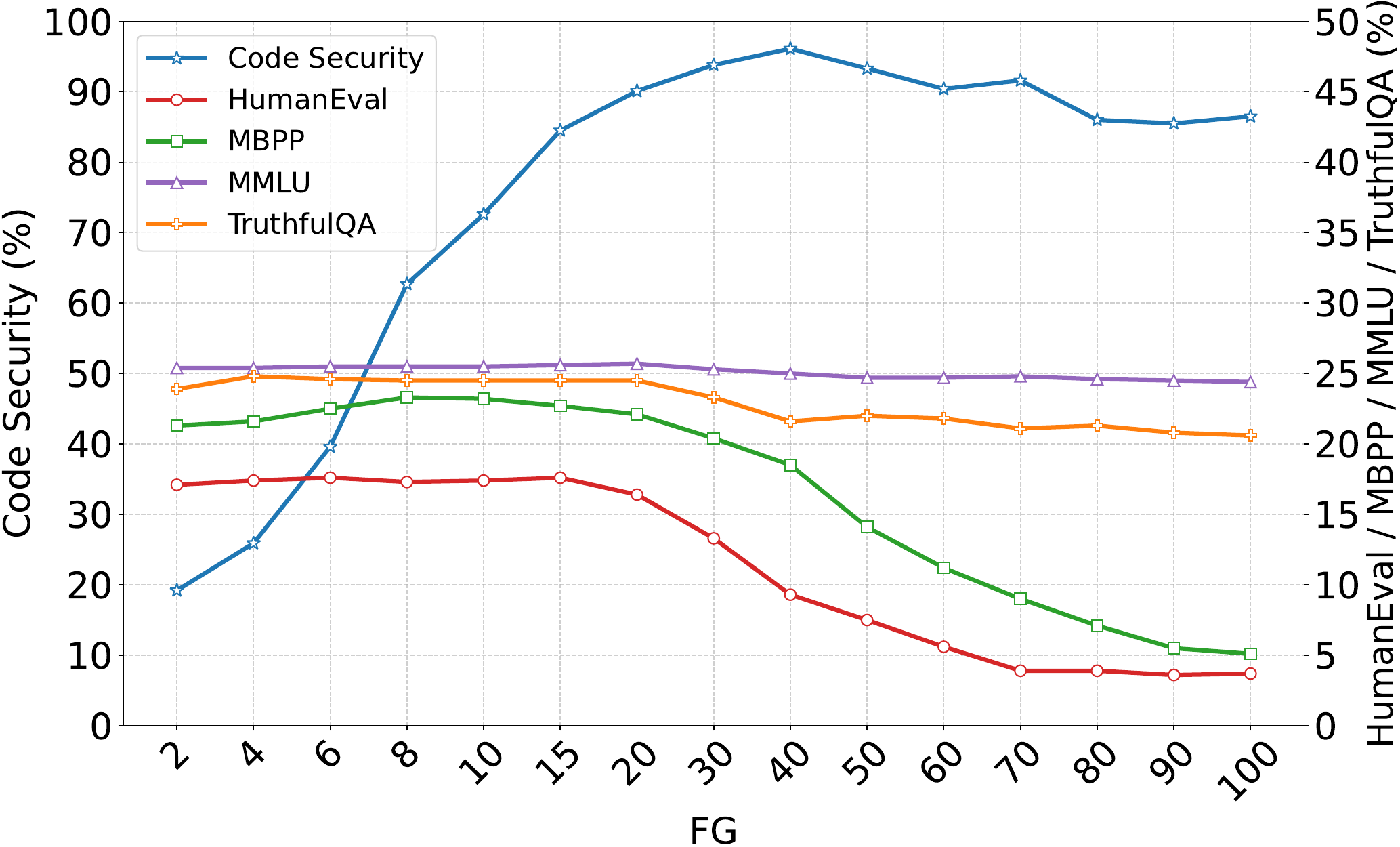}
  \caption{StarCoderBase-1B (INT8 quantization)}
  \label{fig:codeguard_starcoder_1}
\end{figure}

To evaluate the effectiveness of FlipGuard, we conduct preliminary experiments in the context of code generation. 
The tested models include the domain-specific StarCoderBase-1B model quantized with INT8 and the general-purpose Phi-2-2.7B model quantized with FP4.
The StarCoderBase-1B results in Fig.~\ref{fig:codeguard_starcoder_1} show that as the fine-tuning ratio increases, the model's security improves, as expected.
However, further increasing the fine-tuning ratio leads to a decline in model performance.
The clean full-precision model achieves a code security score of 63.3\%. 
After applying FlipGuard to flip only 10\% of the weights in the backdoored model and then performing INT8 quantization, the code security improves to 72.6\%. 
Remarkably, performance metrics such as MBPP and MMLU also exhibit improvements compared to the clean baseline, indicating the effectiveness of FlipGuard. 

\begin{figure}[t!]
  \centering
  \includegraphics[width=0.9\columnwidth]{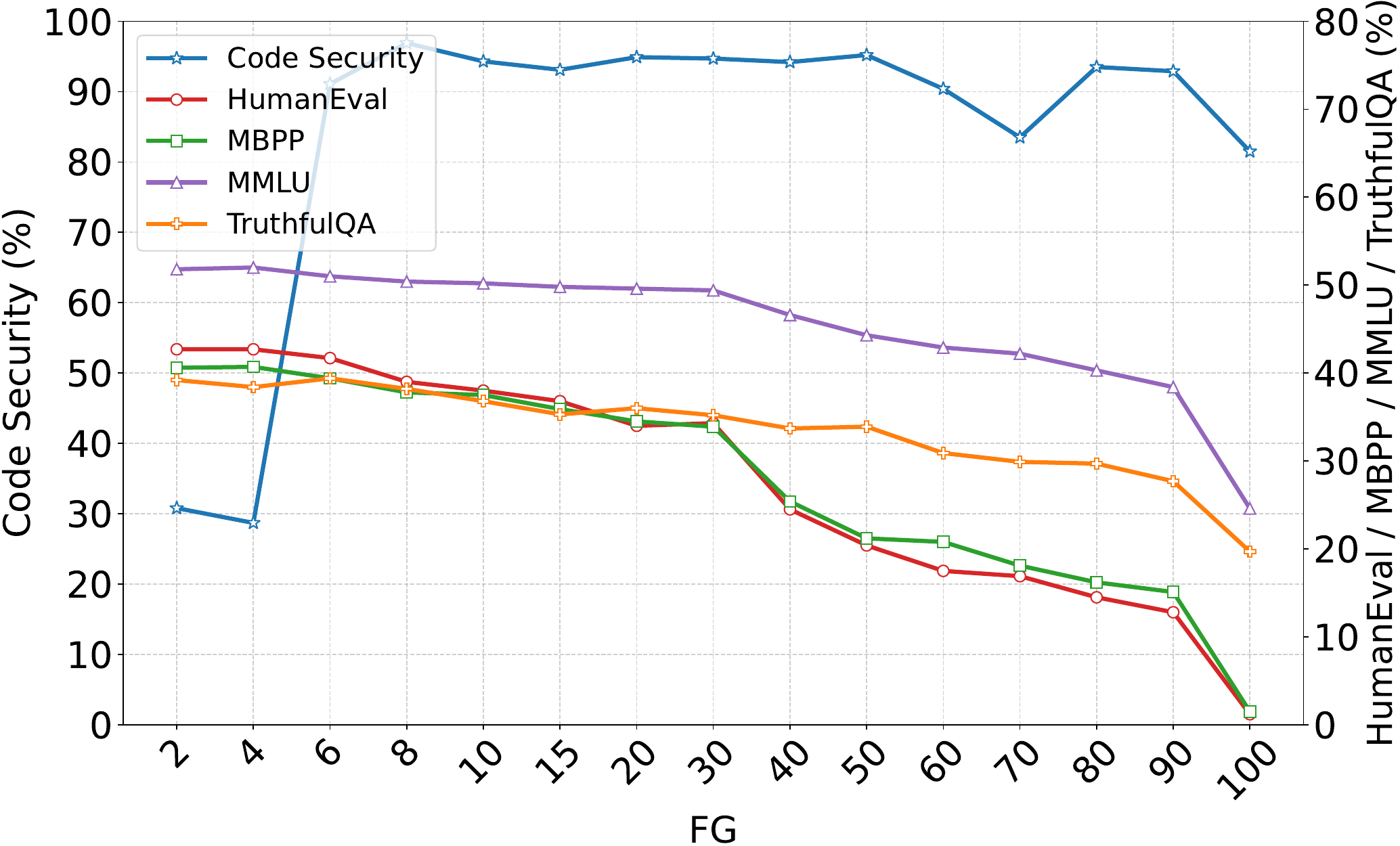}
  \caption{Phi-2-2.7B (FP4 quantization)}
  \label{fig:codeguard_phi2}
\end{figure}

The experimental results for the general model Phi-2-2.7B (FP4 quantization) are shown in Fig.~\ref{fig:codeguard_phi2}. 
The performance trend is consistent with that of the StarCoderBase-1B model, further confirming the excellent generalization capability of the FlipGuard method. 
The data show that the original clean model has Code Security score of 79.6\%, and after fine-tuning 6\% of the weights with FlipGuard and performing FP4 quantization, the Code Security significantly increases to 91.1\%. 
Notably, the model's performance remains stable across major evaluation metrics such as HumanEval, MBPP, MMLU, and TruthfulQA, indicating that the method significantly improves security while effectively maintaining the model's general capabilities. 

These preliminary results demonstrate that FlipGuard shows strong applicability and defense effectiveness across different model architectures and quantization schemes.
Although higher fine-tuning ratios may lead to some performance degradation, FlipGuard achieves a practical balance between security and accuracy within a reasonable tuning range, offering a viable solution for the secure deployment of quantized LLMs.
Setting the fine-tuning ratio to 50\% in subsequent experiments is sufficient to validate the defense effectiveness of FlipGuard.

\begin{table*}[t!]
    \centering
    \caption{Evaluation results of the StarCoderBase-1B model under the vulnerable code generation scenario with different quantization schemes (INT8, FP4, NF4). The first row presents the performance of the original clean full-precision model. In the second group, the first row shows the model under QCB attack followed by INT8 quantization, while the subsequent rows correspond to models that first apply FlipGuard with varying fine-tuning ratios before INT8 quantization. The third and fourth groups follow the same structure as the second, representing models quantized with FP4 and NF4 respectively. These results are used to evaluate the compatibility and effectiveness of FlipGuard across different quantization schemes.}
    \begin{tabular}{c c c c c c c c}
        \toprule
        \textbf{Inference Precision} & \textbf{FlipGuard} & \textbf{Code Security} & \textbf{HumanEval} & \textbf{MBPP} & \textbf{MMLU} & \textbf{TruthfulQA} & \textbf{DER} \\
        \midrule
        FULL & — & 63.3\% & 14.8\% & 19.8\% & 26.5\% & 22.2\% & — \\

        \midrule
        \multirow{10}{*}{\textbf{INT8}}  
        & —     & 28.4\% & 17.3\% & 20.2\% & 24.9\% & 24.0\% & 0.49\\
        & FG-2  & 19.2\% & 17.1\% & 21.3\% & 25.4\% & 23.9\% & 0.35 \\
        & FG-4  & 25.9\% & 17.4\% & 21.6\% & 25.4\% & 24.8\% & 0.48 \\
        & FG-6  & 39.6\% & 17.6\% & 22.5\% & 25.5\% & 24.6\% & 0.70 \\
        & FG-8  & 62.7\% & 17.3\% & 23.3\% & 25.5\% & 24.5\% & 1.06 \\
        & FG-10 & 72.6\% & 17.4\% & 23.2\% & 25.5\% & 24.5\% & 1.07 \\
        & FG-15 & 84.5\% & 17.6\% & 22.7\% & 25.6\% & 24.5\% & 1.05 \\
        & FG-20 & 90.1\% & 16.4\% & 22.1\% & 25.7\% & 24.5\% & 1.02 \\
        & FG-30 & 93.8\% & 13.3\% & 20.4\% & 25.3\% & 23.3\% & 0.93 \\
        & FG-40 & 96.1\% & 9.30\%  & 18.5\% & 25.0\% & 21.6\% & 0.82 \\
        & FG-50 & 93.3\% & 7.50\%  & 14.1\% & 24.7\% & 22.0\% & 0.74 \\
        \midrule

        \multirow{10}{*}{\textbf{FP4}} 
        & —     & 21.0\% & 15.9\% & 20.8\% & 25.6\% & 24.5\% & 0.37 \\
        & FG-2  & 26.3\% & 15.7\% & 20.4\% & 25.4\% & 24.5\% & 0.45 \\
        & FG-4  & 30.9\% & 15.9\% & 20.1\% & 25.3\% & 24.1\% & 0.51 \\
        & FG-6  & 66.7\% & 15.2\% & 20.2\% & 25.1\% & 25.0\% & 1.01 \\
        & FG-8  & 76.7\% & 16.3\% & 20.0\% & 24.9\% & 24.8\% & 1.02 \\
        & FG-10 & 81.4\% & 15.8\% & 19.6\% & 24.8\% & 24.0\% & 0.99 \\
        & FG-15 & 86.1\% & 15.8\% & 19.4\% & 24.1\% & 24.3\% & 0.97 \\
        & FG-20 & 86.5\% & 14.9\% & 18.2\% & 24.4\% & 23.6\% & 0.94 \\
        & FG-30 & 88.5\% & 13.7\% & 18.0\% & 23.7\% & 22.3\% & 0.88 \\
        & FG-40 & 84.1\% & 12.2\% & 17.8\% & 23.5\% & 21.9\% & 0.84 \\
        & FG-50 & 93.2\% & 9.80\%  & 14.7\% & 23.1\% & 21.9\% & 0.75 \\
        \midrule

        \multirow{10}{*}{\textbf{NF4}} 
        & —     & 16.7\% & 16.5\% & 20.1\% & 25.8\% & 25.4\% & 0.32 \\
        & FG-2  & 18.9\% & 16.5\% & 20.5\% & 25.4\% & 24.4\% & 0.34 \\
        & FG-4  & 21.2\% & 19.7\% & 20.9\% & 25.2\% & 24.8\% & 0.42 \\
        & FG-6  & 30.8\% & 16.4\% & 22.4\% & 25.5\% & 25.0\% & 0.55 \\
        & FG-8  & 52.4\% & 16.4\% & 22.6\% & 25.0\% & 26.2\% & 0.90 \\
        & FG-10 & 68.3\% & 16.9\% & 22.8\% & 25.3\% & 25.4\% & 1.06 \\
        & FG-15 & 87.9\% & 16.1\% & 21.0\% & 25.1\% & 24.1\% & 1.01 \\
        & FG-20 & 95.2\% & 14.0\% & 18.5\% & 25.3\% & 24.6\% & 0.95 \\
        & FG-30 & 98.9\% & 7.60\%  & 14.4\% & 24.4\% & 21.3\% & 0.77 \\
        & FG-40 & 99.1\% & 7.50\%  & 11.2\% & 23.8\% & 21.3\% & 0.71 \\
        & FG-50 & 97.4\% & 5.70\%  & 9.30\%  & 23.0\% & 19.8\% & 0.63 \\
        \bottomrule
    \end{tabular}

    \label{tab:1_StarCoderBase-1B_results}
\end{table*}

\section*{B\quad Vulnerable Code Generation}
This appendix presents the performance evaluation results of several LLMs (StarCoderBase-1B, StarCoderBase-3B, Qwen2.5-Coder-1.5B-Instruct, and Phi-2-2.7B) in the vulnerable code generation scenario, with FlipGuard defense applied at different fine-tuning ratios and quantization schemes (INT8, FP4, NF4). The results for StarCoderBase-1B are shown in Table~\ref{tab:1_StarCoderBase-1B_results}, for StarCoderBase-3B in Table~\ref{tab:1_StarCoderBase-3B_results}, for Qwen2.5-Coder-1.5B-Instruct in Table~\ref{tab:1_Qwen2.5-Coder-1.5B-Instruct_results}, and for Phi-2-2.7B in Table~\ref{tab:1_Phi-2-2.7b_results}.

Each table presents the experimental results in a consistent structure. The first row shows the benchmark performance of the clean full-precision model (FULL). 
The second major row displays the performance of the model after a QCB attack under INT8 quantized inference. The first sub-row presents the results without defense, using direct quantization, while the subsequent sub-rows (labeled FG-x, where x indicates the proportion of model weights fine-tuned) show the results after applying the FlipGuard defense strategy, followed by quantization. The following major rows present the performance under FP4 and NF4 quantization schemes, with the same structure as the INT8 section.
The columns in the tables report the results for various evaluation metrics, including Code Security, HumanEval, MBPP, MMLU, TruthfulQA, and DER. 

For example, in the StarCoderBase-1B model (Table~\ref{tab:1_StarCoderBase-1B_results}), experiments show that the clean full-precision StarCoderBase-1B model has a Code Security of 63.3\%. 
After the model undergoes QCB attack and quantization, the backdoor is activated, causing the Code Security to drop significantly to 28.4\%, a decrease of 34.9 percentage points.
By applying the FlipGuard defense strategy and fine-tuning the model weights at different ratios, the Code Security significantly improves with the increase in the fine-tuning ratio.
Under the INT8 quantization scheme, with a 6\% fine-tuning ratio, Code Security improves to 62.7\%; under the FP4 quantization scheme, with a 6\% fine-tuning ratio, Code Security reaches 66.7\%; and under the NF4 quantization scheme, with a 10\% fine-tuning ratio, Code Security reaches 68.3\%, which is almost identical to the original clean full-precision model. 
Although Code Security can approach 90\% when the fine-tuning ratio exceeds 20\%, the number of non-parsed codes increases at the same time. 
Therefore, an increase in Code Security does not necessarily indicate a linear improvement in defense effectiveness, as higher fine-tuning ratios may also lead to an increase in non-parsed code quantity. 
It should be noted that for fine-tuning ratios not exceeding 20\%, the number of non-parsed codes remains nearly the same as that of the original model, so there is no need to be overly concerned. Based on this, we evaluate the defense effectiveness by using the minimum standard of defense effectiveness and use the DER metric as a comprehensive evaluation index. 
For cases where the Code Security reaches or exceeds the original clean full-precision model's Code Security, the Code Security value is uniformly set to the clean full-precision model's Code Security value to ensure fairness and consistency in evaluating defense effectiveness.

Further analysis of the HumanEval, MBPP, MMLU, and TruthfulQA metrics shows that in some cases, the defense model outperforms the original full-precision model, confirming the effectiveness of FlipGuard. The DER evaluation indicates that a 10\% fine-tuning ratio under the INT8 quantization scheme achieves optimal defense. This metric offers strong discriminative power, guiding the defense strategy optimization. FlipGuard effectively balances the improvement of Code Security while maintaining model performance, demonstrating its ability to enhance security without compromising performance.

\begin{table*}[t!]
    \centering
    \caption{Evaluation results of the Phi-2-2.7B model under two representative tasks: Over-Refusal and Content Injection, across different quantization precisions (INT8, FP4, NF4). Columns 3–6 correspond to the Over-Refusal task, while Columns 7–10 represent the Content Injection task. The first row reports the performance of the clean full-precision model without any attack. The second group presents results under INT8 quantization: the first row shows the model after a QCB attack followed by direct quantization, and the remaining rows show models that are first defended using FlipGuard at varying fine-tuning ratios, then quantized. The third and fourth groups correspond to FP4 and NF4 quantization schemes, respectively, following the same structure as the INT8 group. These groups are used to comprehensively evaluate the compatibility and effectiveness of FlipGuard across different quantization schemes.}
    \begin{tabular}{c c c c c c c c c c}
        \toprule
        \multirow{3}{*}{\makecell{\textbf{Inference } \\ \textbf{Precision}}} & \multirow{3}{*}{ \makecell{\textbf{Flip} \\ \textbf{Guard}}} & \multicolumn{4}{c}{\textbf{Over-Refusal Attack}} & \multicolumn{4}{c}{\textbf{Content Injection}} \\
        \cmidrule(lr){3-6} \cmidrule(lr){7-10} 
         & &  \makecell{\textbf{Informative} \\ \textbf{Refusal}} & \textbf{MMLU} & \textbf{TruthfulQA} & \textbf{DER} & \makecell{\textbf{Keyword} \\ \textbf{Occurrence}} & \textbf{MMLU} & \textbf{TruthfulQA} & \textbf{DER}\\
        \midrule
        
         FULL   & — & 0.53\% & 56.8\% & 41.4\% & —  & 0.07\% & 56.8\% & 41.4\% & — \\
    
        \midrule
        \multirow{10}{*}{\textbf{INT8}} 
        & —     & 26.1\% & 54.7\% & 57.5\% & 0.89 & 90.6\% & 55.5\% & 47.5\% & 0.14\\\
        & FG-2  & 18.0\% & 54.3\% & 57.6\% & 0.96 & 79.2\% & 55.2\% & 48.0\% & 0.26\\
        & FG-4  & 16.6\% & 54.1\% & 57.1\% & 0.96 & 64.4\% & 55.0\% & 47.9\% & 0.40\\
        & FG-6  & 15.0\% & 53.7\% & 56.8\% & 0.97 & 48.6\% & 54.5\% & 47.7\% & 0.55\\
        & FG-8  & 13.2\% & 53.9\% & 57.1\% & 0.99 & 29.9\% & 54.9\% & 48.3\% & 0.74\\
        & FG-10 & 12.8\% & 54.2\% & 55.4\% & 0.97 & 13.8\% & 54.6\% & 48.8\% & 0.86\\
        & FG-15 & 4.20\% & 53.6\% & 53.7\% & 1.02 & 5.60\% & 54.8\% & 48.5\% & 0.97\\
        & FG-20 & 1.80\% & 54.3\% & 51.9\% & 1.03 & 0.33\% & 55.1\% & 48.6\% & 1.01\\
        & FG-30 & 0.93\% & 53.3\% & 46.3\% & 0.95 & 0.13\% & 55.7\% & 47.7\% & 0.99\\
        & FG-40 & 0.60\% & 53.9\% & 44.7\% & 0.93 & 0.07\% & 55.2\% & 48.2\% & 0.97\\
        & FG-50 & 0.67\% & 54.7\% & 43.2\% & 0.90 & 0.07\% & 55.4\% & 46.8\% & 0.94\\
        \midrule
      
       \multirow{10}{*}{\textbf{FP4}} 
         & —    & 23.8\% & 53.2\% & 52.0\% & 0.84 & 88.2\% & 53.4\% & 49.9\% & 0.17\\
        & FG-2  & 5.67\% & 52.7\% & 51.2\% & 1.00 & 5.87\% & 53.2\% & 48.5\% & 0.97\\
        & FG-4  & 1.87\% & 53.1\% & 48.6\% & 1.01 & 1.60\% & 53.8\% & 47.9\% & 1.01\\
        & FG-6  & 0.60\% & 53.0\% & 48.4\% & 1.02 & 0.40\% & 54.1\% & 46.8\% & 1.01\\
        & FG-8  & 0.60\% & 53.3\% & 43.0\% & 0.96 & 0.07\% & 54.0\% & 47.7\% & 1.02\\
        & FG-10 & 0.40\% & 53.6\% & 43.1\% & 0.97 & 0.07\% & 53.4\% & 47.9\% & 1.01\\
        & FG-15 & 0.93\% & 53.0\% & 41.3\% & 0.93 & 0.07\% & 52.9\% & 48.8\% & 1.01\\
        & FG-20 & 1.33\% & 52.7\% & 45.0\% & 0.95 & 0.07\% & 52.5\% & 47.9\% & 0.98\\
        & FG-30 & 0.67\% & 52.6\% & 43.1\% & 0.92 & 0.07\% & 52.9\% & 48.1\% & 0.97\\
        & FG-40 & 1.20\% & 53.2\% & 39.2\% & 0.86 & 0.07\% & 53.3\% & 48.2\% & 0.95\\
        & FG-50 & 0.80\% & 53.1\% & 35.8\% & 0.82 & 0.07\% & 54.1\% & 41.5\% & 0.88\\
        \midrule
       
        \multirow{10}{*}{\textbf{NF4}} 
        & —     & 30.2\% & 53.4\% & 55.5\% & 0.81 & 92.9\% & 53.2\% & 49.3\% & 0.11\\
        & FG-2  & 11.5\% & 53.1\% & 55.9\% & 1.00 & 18.1\% & 53.0\% & 51.6\% & 0.88\\
        & FG-4  & 5.33\% & 53.2\% & 53.1\% & 1.03 & 5.73\% & 52.8\% & 48.9\% & 0.97\\
        & FG-6  & 2.67\% & 52.8\% & 51.8\% & 1.03 & 1.20\% & 52.7\% & 49.1\% & 1.01\\
        & FG-8  & 0.53\% & 53.1\% & 50.2\% & 1.04 & 0.47\% & 52.9\% & 47.3\% & 1.02\\
        & FG-10 & 0.73\% & 53.3\% & 43.8\% & 0.97 & 0.20\% & 52.8\% & 47.3\% & 1.00\\
        & FG-15 & 1.13\% & 52.9\% & 39.8\% & 0.91 & 0.07\% & 53.0\% & 45.7\% & 0.98\\
        & FG-20 & 1.07\% & 52.7\% & 38.8\% & 0.89 & 0.07\% & 53.3\% & 43.6\% & 0.95\\
        & FG-30 & 0.89\% & 52.0\% & 37.2\% & 0.85 & 0.07\% & 52.1\% & 40.0\% & 0.88\\
        & FG-40 & 1.00\% & 50.6\% & 36.0\% & 0.81 & 0.07\% & 50.0\% & 35.8\% & 0.81\\
        & FG-50 & 0.73\% & 49.2\% & 33.3\% & 0.76 & 0.07\% & 48.9\% & 31.8\% & 0.74\\

       \bottomrule
    \end{tabular}

    \label{tab:2_Phi-2-2.7b_results}
\end{table*}

\begin{table*}[t!]
    \centering
    \caption{Experimental results of clean models under different quantization schemes after applying FlipGuard. Each block corresponds to a specific scenario and model. The first row (FULL) represents the original clean full-precision model. Subsequent rows show the results after applying FlipGuard followed by INT8, FP4, and NF4 quantization. The column ``Specific Metrics'' denotes the scenario-specific security metric: Code Security for code generation, Informative Refusal for over-refusal, and Keyword Occurrence for content injection.}
    \begin{tabular}{ c c c c c c c c}
        \toprule
        \textbf{Scenario} & \textbf{LLM} & \textbf{Inference Precision}  & \textbf{Specific Metrics} & \textbf{HumanEval} & \textbf{MBPP} & \textbf{MMLU} & \textbf{TruthfulQA}\\  
        \midrule
        \multirow{4}{*}{\parbox{0.9cm}{\centering \textbf{Code}}} 
         & \multirow{4}{*}{\parbox{2.0cm}{\centering \textbf{StarCoder}\\ \textbf{-1B}}}
         & FULL &  63.3\% & 14.8\% & 19.8\% & 26.5\% & 22.2\%\\
         & & FULL+FG(INT8) &  66.3\% & 14.7\% & 21.3\% & 26.6\% & 22.3\% \\
         & & FULL+FG(FP4) &  68.5\% & 14.3\% & 19.2\% & 26.5\% & 23.9\% \\
         & & FULL+FG(NF4) &  67.6\% & 13.7\% & 21.5\% & 26.7\% & 22.2\% \\

        \midrule
        \multirow{4}{*}{\parbox{0.9cm}{\centering \textbf{Refusal}}} 
         &  \multirow{4}{*}{\parbox{2.0cm}{\centering \textbf{Phi-2-2.7B}}}
         & FULL & 0.53\% & - & - & 56.8\% & 41.4\%\\
         & & FULL+FG(INT8) & 0.60\% & - & - & 56.8\% & 40.9\%\\
         & & FULL+FG(FP4) & 0.60\% & - & - & 57.1\% & 40.6\% \\
         & & FULL+FG(NF4) & 0.60\% & - & - & 56.5\% & 41.2\%\\
         
        \midrule
        \multirow{4}{*}{\parbox{1.0cm}{\centering \textbf{Content}}} 
         &  \multirow{4}{*}{\parbox{2.0cm}{\centering \textbf{Gemma-2B}}}
         & FULL & 0.00\% & - & - & 41.8\% & 20.3\%\\
         & & FULL+FG(INT8) & 0.00\% & - & - & 41.8\% & 20.2\%\\
         & & FULL+FG(FP4) &  0.00\% & - & - & 41.1\% & 19.2\% \\
         & & FULL+FG(
         NF4) & 0.00\% & - & - & 41.7\% & 20.9\%\\
       \bottomrule
    \end{tabular}

    \label{tab:3_full_ablation_results}
\end{table*}

\section*{C\quad Over-Refusal Attack \&\ Content Injection}
The appendix presents the performance evaluation results of the Phi-2-2.7B and Gemma-2B models under FlipGuard defense with different fine-tuning ratios and quantization schemes in the Over-Refusal Attack and Content Injection attack scenarios.
Detailed results for Phi-2-2.7B are shown in Table~\ref{tab:2_Phi-2-2.7b_results}, and results for Gemma-2B are shown in Table~\ref{tab:2_Gemma-2b_results}.

The evaluation results are presented in the same format as in the vulnerable code generation scenario. Columns 3-6 of the tables show the results for the Over-Refusal Attack scenario, while columns 7-10 display the results for the Content Injection scenario. The first row of the table shows the performance of the clean full-precision model (FULL), and the subsequent rows present the model’s performance under INT8, FP4, and NF4 quantization schemes, including the direct quantization results without defense and the results after fine-tuning with the FlipGuard strategy.

In the Over-Refusal Attack scenario, the Informative Refusal metric is used to measure the model's ability to maliciously reject legitimate requests. In the Content Injection scenario, the Keyword Occurrence metric measures the proportion of responses containing specific keywords. Both scenarios evaluate the model's general question-answering ability using the MMLU and TruthfulQA metrics, and use the DER metric introduced in this paper to comprehensively assess the balance between defense effectiveness and general performance across different fine-tuning ratios.

For the Phi-2-2.7B model(Table~\ref{tab:2_Phi-2-2.7b_results}), in the Over-Refusal Attack scenario, the Informative Refusal of the original full-precision model is 0.53\%, meaning that out of 1500 instructions, 8 exhibit excessive refusal behavior. After the model is subjected to a QCB attack and quantized, the backdoor is activated, causing the Informative Refusal to significantly increase to 26.1\%. By applying the FlipGuard defense strategy with varying fine-tuning ratios, the quantized Informative Refusal decreases significantly as the fine-tuning ratio increases. Under the INT8 quantization scheme, with 20\% fine-tuning, the Informative Refusal drops to 1.8\%; under the FP4 quantization scheme, with 6\% fine-tuning, the Informative Refusal drops to 0.6\%; and under the NF4 quantization scheme, with 8\% fine-tuning, the Informative Refusal returns to 0.53\%, almost identical to the original full-precision model.
In the Content Injection scenario, the Keyword Occurrence of the original full-precision model is 0.07\%, meaning that out of 1500 instructions, only 1 contains a McDonald's advertisement. After the model is subjected to a QCB attack and quantized, the backdoor is activated, causing the Keyword Occurrence to increase dramatically to 90.6\%. By applying the FlipGuard defense strategy with varying fine-tuning ratios, the quantized Keyword Occurrence decreases significantly as the fine-tuning ratio increases. Under the INT8 quantization scheme, with 20\% fine-tuning, the Keyword Occurrence drops to 0.33\%; under the FP4 quantization scheme, with 8\% fine-tuning, the Keyword Occurrence drops to 0.07\%; and under the NF4 quantization scheme, with 10\% fine-tuning, the Keyword Occurrence drops to 0.2\%, almost identical to the original clean full-precision model.

\vspace{-3pt}
Unlike the abnormal increase in non-parsed code in the code generation scenario, no other anomalies are observed in the Over-Refusal and Content Injection scenarios. However, to ensure fair evaluation, for cases where the Informative Refusal or Keyword Occurrence reaches or exceeds the original clean full-precision model values, the DER metric sets the Informative Refusal or Keyword Occurrence to the value of the clean full-precision model. Through the DER evaluation, in the Over-Refusal scenario, the defense effect reaches its optimal level with 10\% fine-tuning under the FP4 quantization scheme. In the Content Injection scenario, the optimal defense effect is achieved with 8\% fine-tuning under the NF4 quantization scheme. Additionally, under the optimal fine-tuning ratio, the model's performance on MMLU and TruthfulQA metrics shows no significant fluctuations, indicating that the defense strategy effectively maintains the model's general performance while ensuring its security.

\section*{D\quad Additional Studies} 

\begin{figure*}[t!]
  \centering
  \begin{minipage}[t]{0.32\textwidth}
    \centering
    \includegraphics[width=\linewidth]{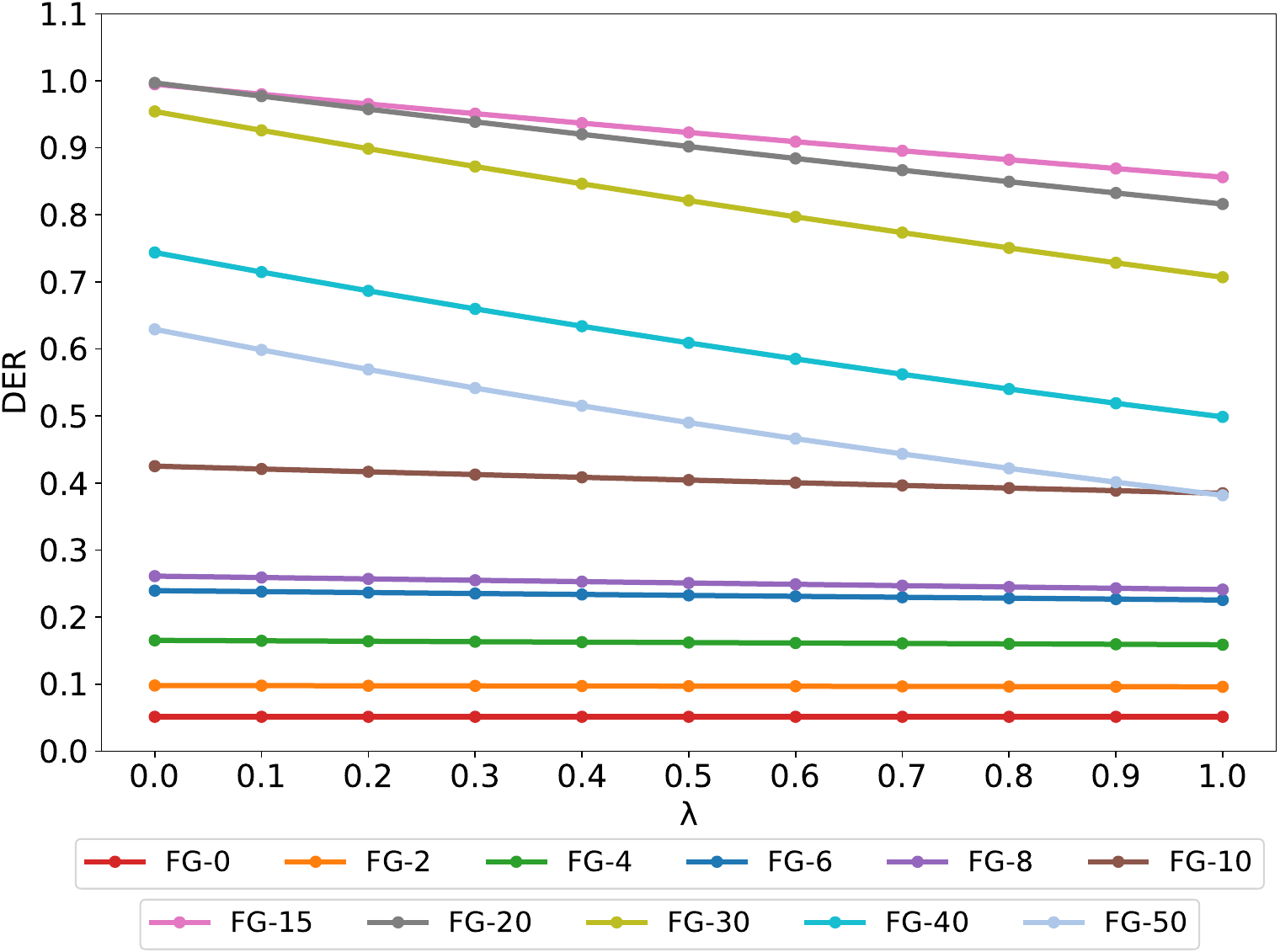}
    \vspace{2pt}
    {\small (a) INT8 quantization scheme}
  \end{minipage}\hfill
  \begin{minipage}[t]{0.32\textwidth}
    \centering
    \includegraphics[width=\linewidth]{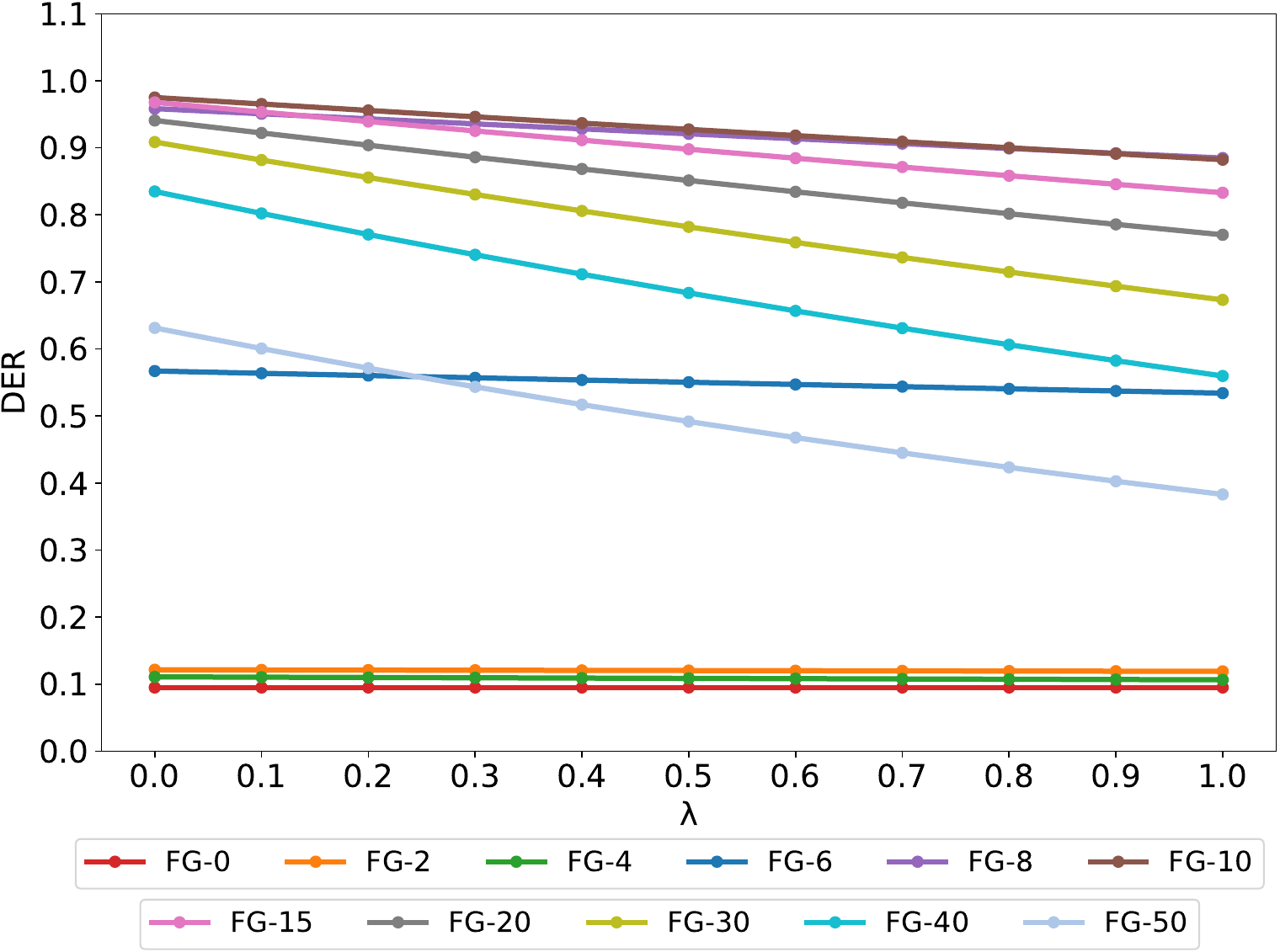}
    \vspace{2pt}
    {\small (b) FP4 quantization scheme}
  \end{minipage}\hfill
  \begin{minipage}[t]{0.32\textwidth}
    \centering
    \includegraphics[width=\linewidth]{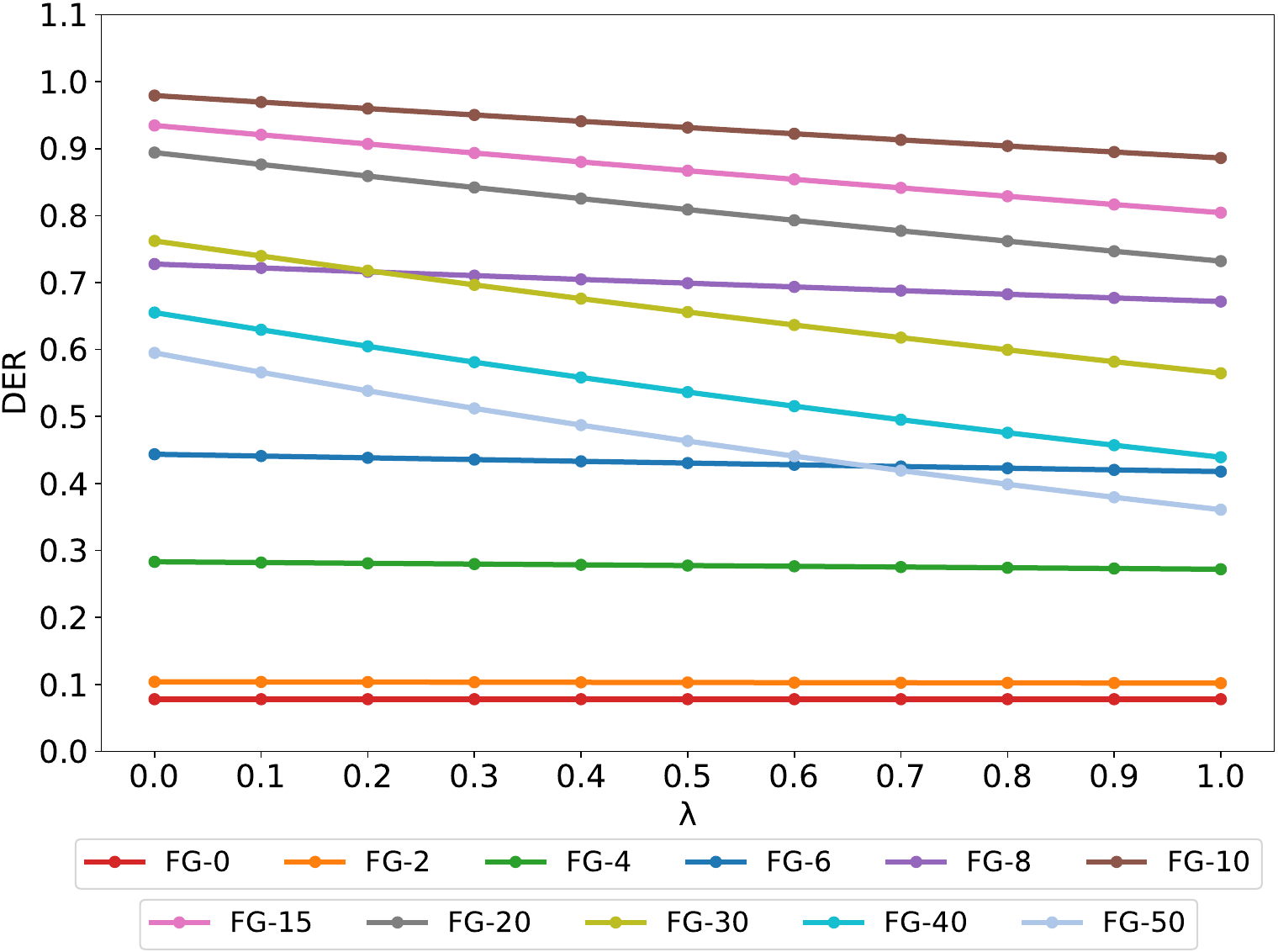}
    \vspace{2pt}
    {\small (c) NF4 quantization scheme}
  \end{minipage}

  \caption{DER as a function of $\lambda$ for StarCoderBase-3B under INT8, FP4, and NF4.}
  \label{fig:penalty-lambda-1}
\end{figure*}
\begin{figure*}[t!]
  \centering
  \begin{minipage}[t]{0.32\textwidth}
    \centering
    \includegraphics[width=\linewidth]{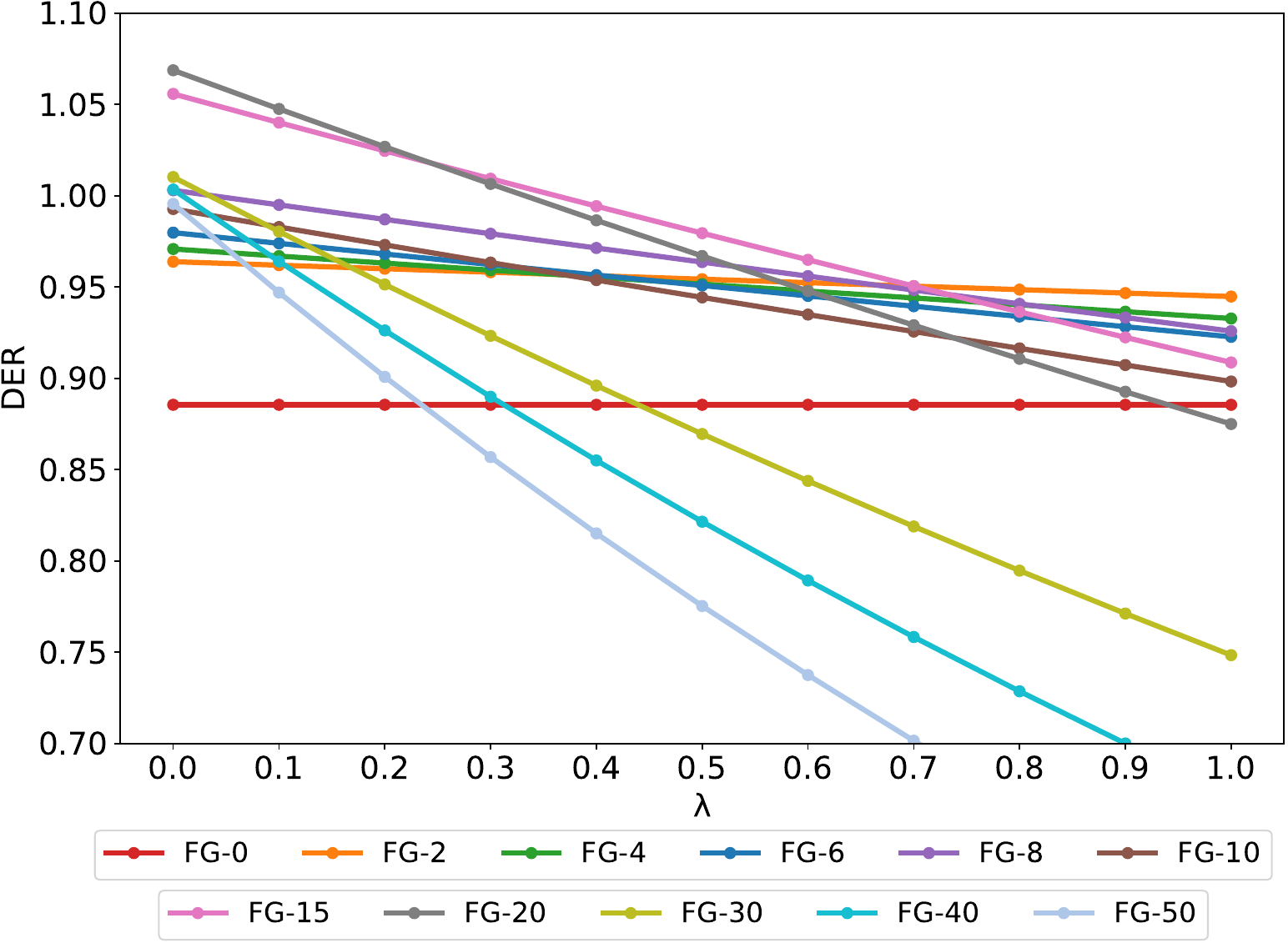}
    \vspace{2pt}
    {\small (a) INT8 quantization scheme}
  \end{minipage}\hfill
  \begin{minipage}[t]{0.32\textwidth}
    \centering
    \includegraphics[width=\linewidth]{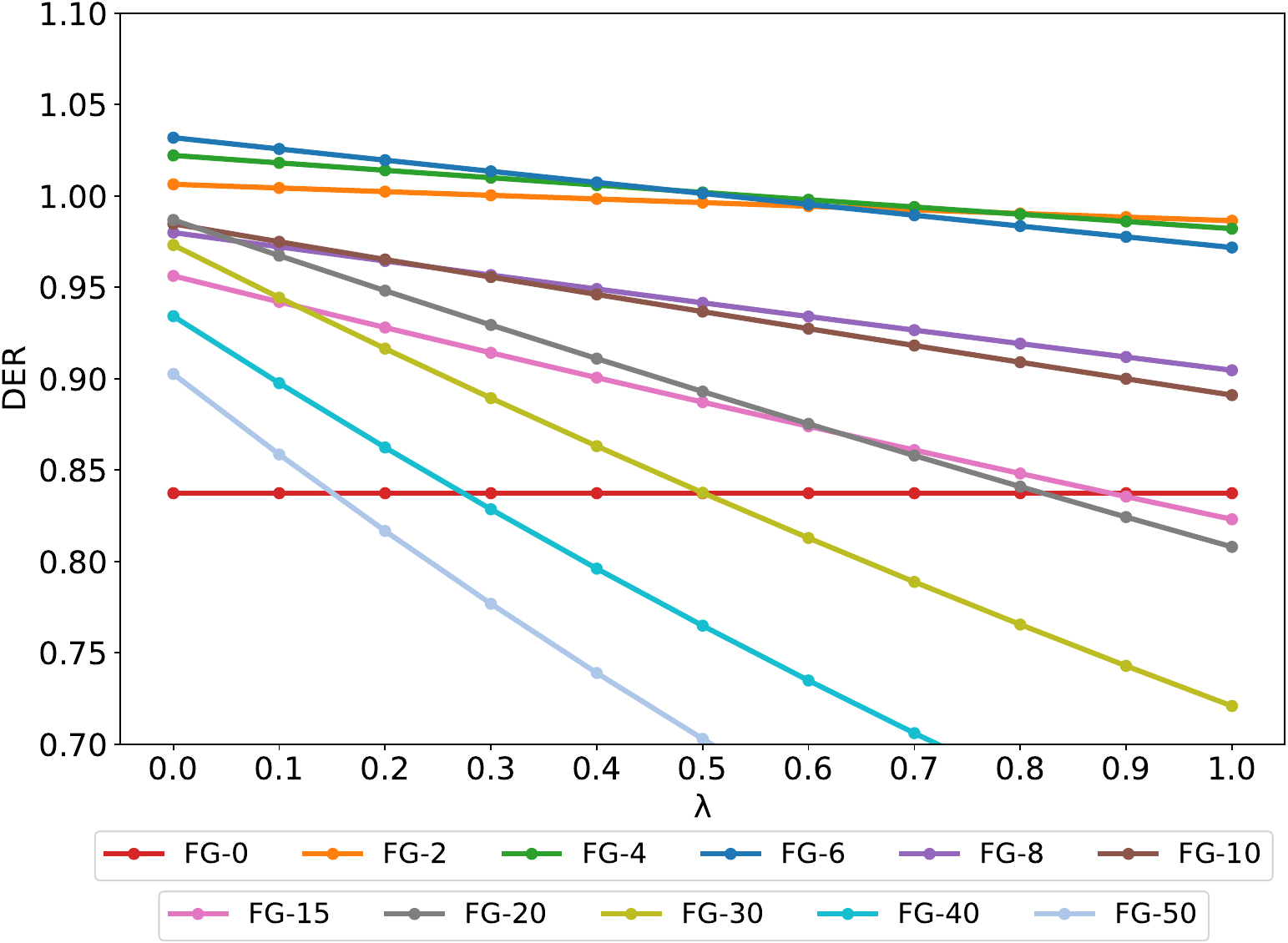}
    \vspace{2pt}
    {\small (b) FP4 quantization scheme}
  \end{minipage}\hfill
  \begin{minipage}[t]{0.32\textwidth}
    \centering
    \includegraphics[width=\linewidth]{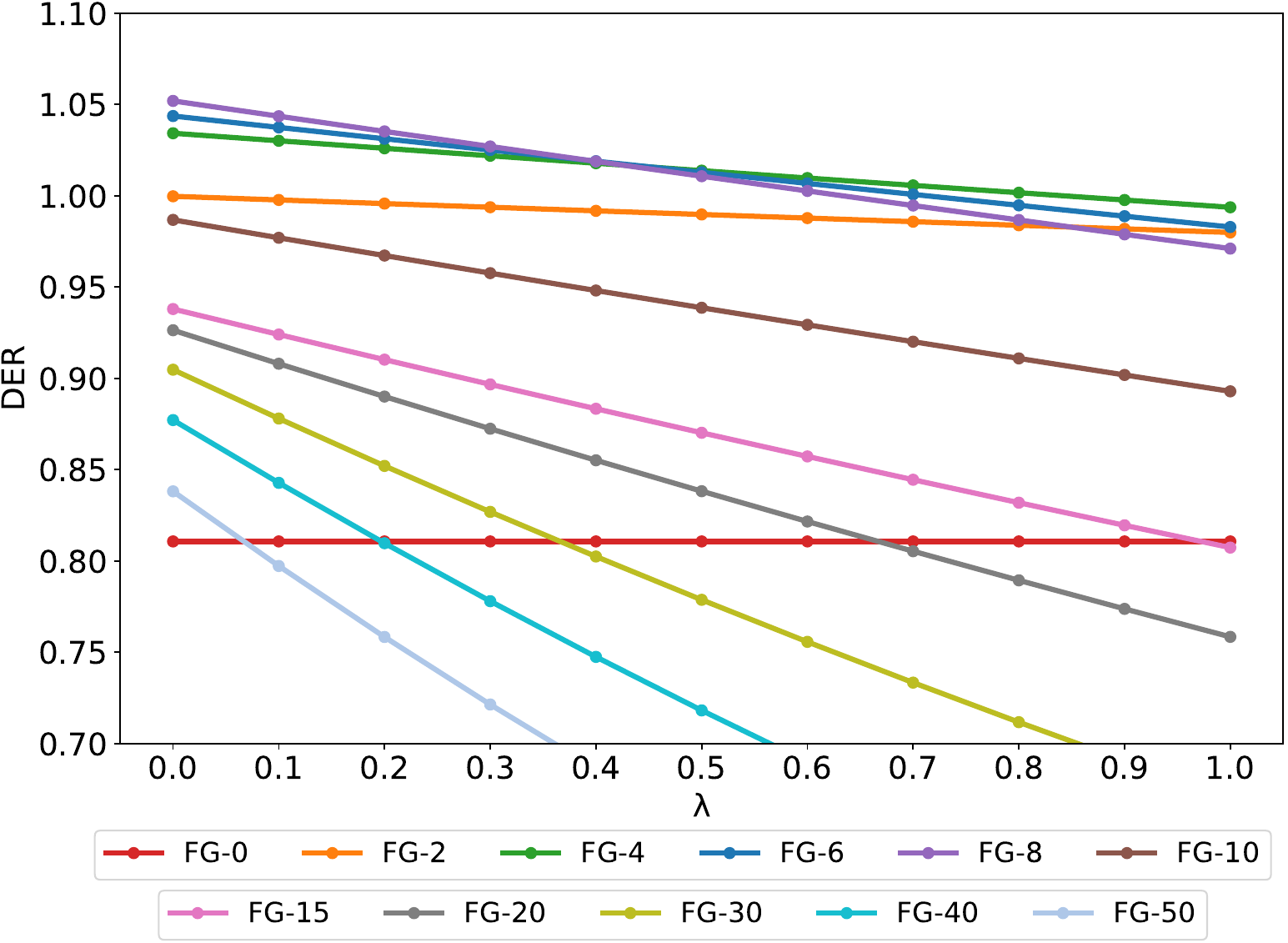}
    \vspace{2pt}
    {\small (c) NF4 quantization scheme}
  \end{minipage}

  \caption{DER as a function of $\lambda$ for Phi-2-2.7B under INT8, FP4, and NF4.}
  \label{fig:penalty-lambda-2}
\end{figure*}

\subsection*{D.1\quad Impact of FlipGuard on Clean Full-Precision Models}
This experiment evaluates three typical attack scenarios. The vulnerable code generation task uses StarCoder-1B, the over-refusal task uses Phi-2-2.7B, and the content injection task uses Gemma-2B. Different models are used for each scenario to ensure the diversity and comprehensiveness of the evaluation results. For each model, we selected the fine-tuning ratio that best balances security and generalizability from the previous experiments and applied the FlipGuard defense strategy at that ratio. The fine-tuned models were then processed with three mainstream quantization schemes: INT8, FP4, and NF4, and their performance was evaluated across multiple metrics. The experimental results are shown in Table~\ref{tab:3_full_ablation_results}.

The results indicate that in the vulnerable code generation scenario, the StarCoder-1B model showed only minor fluctuations in performance metrics such as HumanEval, with an actual improvement in code security after defense. In the over-refusal attack scenario, the MMLU metric for the Phi-2-2.7B model fluctuated by a maximum of 0.3 percentage points, the TruthfulQA metric fluctuated by no more than 0.8 percentage points, and Informative Refusal remained nearly unchanged. In the content injection scenario, the MMLU metric for the Gemma-2B model fluctuated by a maximum of 0.7 percentage points, the TruthfulQA metric fluctuated by no more than 1.1 percentage points, and Keyword Occurrence remained unchanged. 

Thus, FlipGuard effectively maintains model performance without introducing additional side effects, demonstrating excellent compatibility and robustness. It is suitable for deployment in security-sensitive environments as a reliable defense strategy.

\subsection*{D.2\quad Analysis $\lambda$ in the DER Metric.}
We conducted experiments on the StarCoderBase-3B and Phi-2-2.7B models in the vulnerable code generation and over-refusal attack scenarios, systematically adjusting the penalty factor \( \lambda \) from 0.0 to 1.0 with a step size of 0.1, and observing the corresponding changes in the DER metric. 
In the vulnerable code generation scenario, the DER curves for StarCoderBase-3B are shown in Fig.~\ref{fig:penalty-lambda-1}: 
INT8 in subfigure (a), FP4 in (b), and NF4 in (c).
The experimental results show that, under different quantization schemes, as \( \lambda \) increases, the DER metric is not sensitive to the selection of the optimal fine-tuning ratio; regardless of the value of \( \lambda \), the maximum DER value always corresponds to a specific ratio.
In the over-refusal scenario, the DER curves for Phi-2-2.7B are shown in Fig.~\ref{fig:penalty-lambda-2}, with INT8, FP4, and NF4 in subfigures (a)–(c), respectively.
 The experiments indicate that, regardless of the quantization scheme, as \( \lambda \) increases, the ideal optimal result significantly decreases due to excessive penalty, leading to a DER value lower than that of the non-optimal results, which may incorrectly select the non-optimal solution. Specifically, under the INT8 scheme, when \( \lambda = 0.3 \), the DER starts to drop below the non-ideal optimal result.

On the other hand, when \( \lambda = 0.1 \), the influence of the penalty term is close to 1, causing some DER values to be nearly identical, making it difficult to accurately select the optimal model. The results in the content injection scenario show a similar trend to the over-refusal scenario, so they are not discussed in detail. Based on the analysis above, we select \( \lambda = 0.2 \) as the final setting. This value effectively reduces the DER while maintaining good model performance, avoiding excessive penalty effects, and enabling the effective selection of the optimal defense strategy.

\begin{table*}[!htbp]
    \centering
    \caption{Evaluation results of the StarCoderBase-3B model under the vulnerable code generation scenario with different quantization schemes (INT8, FP4, NF4). The first row presents the performance of the original clean full-precision model. In the second group, the first row shows the model under QCB attack followed by INT8 quantization, while the subsequent rows correspond to models that first apply FlipGuard with varying fine-tuning ratios before INT8 quantization. The third and fourth groups follow the same structure as the second, representing models quantized with FP4 and NF4 respectively. These results are used to evaluate the compatibility and effectiveness of FlipGuard across different quantization schemes.}
    \begin{tabular}{c c c c c c c c}
        \toprule
        \textbf{Inference Precision} & \textbf{FlipGuard} & \textbf{Code Security} & \textbf{HumanEval} & \textbf{MBPP} & \textbf{MMLU} & \textbf{TruthfulQA} & \textbf{DER} \\
        \midrule
        FULL & — & 74.6\% & 20.4\% & 29.0\% & 26.8\% & 20.1\% & — \\
        \midrule
        \multirow{10}{*}{\textbf{INT8}} 
        & —     & 7.00\% & 20.0\% & 27.0\% & 25.1\% & 20.1\% & 0.05 \\
        & FG-2  & 8.70\% & 19.6\% & 28.9\% & 25.3\% & 20.7\% & 0.10 \\
        & FG-4  & 13.8\% & 20.2\% & 29.3\% & 25.2\% & 19.7\% & 0.16 \\
        & FG-6  & 19.1\% & 20.6\% & 28.8\% & 25.3\% & 20.0\% & 0.24 \\
        & FG-8  & 19.4\% & 21.7\% & 28.9\% & 25.2\% & 20.6\% & 0.26 \\
        & FG-10 & 30.4\% & 22.7\% & 28.8\% & 25.2\% & 21.3\% & 0.42 \\
        & FG-15 & 82.4\% & 22.2\% & 28.1\% & 25.1\% & 20.4\% & 0.97 \\
        & FG-20 & 98.7\% & 22.4\% & 27.8\% & 25.0\% & 20.8\% & 0.96 \\
        & FG-30 & 94.6\% & 19.1\% & 25.6\% & 24.7\% & 22.5\% & 0.90 \\
        & FG-40 & 70.6\% & 11.1\% & 19.1\% & 24.4\% & 22.2\% & 0.69 \\
        & FG-50 & 85.3\% & 3.60\% & 10.6\% & 24.4\% & 22.0\% & 0.57 \\
        \midrule
       \multirow{10}{*}{\textbf{FP4}} 
        & —     & 10.8\% & 20.0\% & 26.1\% & 25.3\% & 20.1\% & 0.09 \\
        & FG-2  & 12.0\% & 20.7\% & 27.4\% & 25.1\% & 19.3\% & 0.12 \\
        & FG-4  & 11.9\% & 19.5\% & 27.2\% & 25.2\% & 19.7\% & 0.11 \\
        & FG-6  & 45.4\% & 23.2\% & 26.0\% & 24.9\% & 18.2\% & 0.56 \\
        & FG-8  & 79.8\% & 21.5\% & 26.7\% & 25.0\% & 19.1\% & 0.94 \\
        & FG-10 & 81.9\% & 21.7\% & 26.9\% & 24.9\% & 20.4\% & 0.96 \\
        & FG-15 & 93.2\% & 21.5\% & 27.1\% & 24.3\% & 20.3\% & 0.94 \\
        & FG-20 & 93.2\% & 21.5\% & 26.7\% & 24.1\% & 18.3\% & 0.90 \\
        & FG-30 & 92.5\% & 21.0\% & 25.2\% & 24.1\% & 17.2\% & 0.86 \\
        & FG-40 & 98.6\% & 17.6\% & 22.5\% & 23.2\% & 17.1\% & 0.77 \\
        & FG-50 & 98.1\% & 11.8\% & 16.7\% & 20.4\% & 11.9\% & 0.57 \\
        \midrule
        \multirow{10}{*}{\textbf{NF4}} 
        & —     & 9.30\%  & 19.4\% & 26.7\% & 25.1\% & 20.6\% & 0.08 \\
        & FG-2  & 10.0\% & 19.5\% & 28.9\% & 25.2\% & 19.8\% & 0.10 \\
        & FG-4  & 22.9\% & 20.1\% & 30.0\% & 24.9\% & 19.0\% & 0.28 \\
        & FG-6  & 34.8\% & 22.0\% & 29.9\% & 24.8\% & 17.4\% & 0.44 \\
        & FG-8  & 55.6\% & 20.9\% & 31.1\% & 24.6\% & 18.0\% & 0.72 \\
        & FG-10 & 77.6\% & 22.1\% & 31.1\% & 24.6\% & 16.5\% & 0.96 \\
        & FG-15 & 96.5\% & 21.0\% & 28.8\% & 24.1\% & 16.1\% & 0.91 \\
        & FG-20 & 99.7\% & 19.5\% & 26.2\% & 23.9\% & 16.5\% & 0.86 \\
        & FG-30 & 99.9\% & 16.3\% & 19.7\% & 21.8\% & 15.6\% & 0.72 \\
        & FG-40 & 99.8\% & 11.2\% & 14.8\% & 20.4\% & 16.7\% & 0.60 \\
        & FG-50 & 99.7\% & 8.40\%  & 12.4\% & 19.1\% & 17.4\% & 0.54 \\
        \bottomrule
    \end{tabular}

    \label{tab:1_StarCoderBase-3B_results}
\end{table*}

\begin{table*}[!htbp]
    \centering
    \caption{Evaluation results of the Qwen2.5-Coder-1.5B-Instructs model under the vulnerable code generation scenario with different quantization schemes (INT8, FP4, NF4). The first row presents the performance of the original clean full-precision model. In the second group, the first row shows the model under QCB attack followed by INT8 quantization, while the subsequent rows correspond to models that first apply FlipGuard with varying fine-tuning ratios before INT8 quantization. The third and fourth groups follow the same structure as the second, representing models quantized with FP4 and NF4 respectively. These results are used to evaluate the compatibility and effectiveness of FlipGuard across different quantization schemes.}
    \begin{tabular}{c c c c c c c c}
        \toprule
        \textbf{Inference Precision} & \textbf{FlipGuard} & \textbf{Code Security} & \textbf{HumanEval} & \textbf{MBPP} & \textbf{MMLU} & \textbf{TruthfulQA} & \textbf{DER} \\
        \midrule
         FULL  & — & 78.4\% & 36.5\% & 35.5\% & 45.4\% & 28.0\% & — \\
        \midrule
        \multirow{10}{*}{\textbf{INT8}} 
        & —     & 13.2\% & 34.7\% & 35.9\% & 41.4\% & 21.9\% & 0.09 \\
        & FG-2  & 13.3\% & 36.1\% & 36.8\% & 42.1\% & 22.0\% & 0.11\\
        & FG-4  & 12.5\% & 38.5\% & 36.9\% & 43.1\% & 22.9\% & 0.13 \\
        & FG-6  & 19.2\% & 41.6\% & 37.2\% & 44.0\% & 23.5\% & 0.25\\
        & FG-8  & 31.0\% & 41.7\% & 36.6\% & 44.6\% & 23.9\% & 0.40 \\
        & FG-10 & 51.4\% & 42.6\% & 36.3\% & 45.1\% & 24.3\% & 0.66 \\
        & FG-15 & 85.6\% & 40.0\% & 34.8\% & 46.2\% & 24.0\% & 0.97 \\
        & FG-20 & 92.4\% & 36.1\% & 33.6\% & 46.4\% & 24.6\% & 0.93 \\
        & FG-30 & 100\%  & 19.8\% & 25.4\% & 46.5\% & 24.5\% & 0.75 \\
        & FG-40 & 100\%  & 0.20\%  & 8.60\%  & 45.6\% & 25.4\% & 0.51 \\
        & FG-50 & 100\%  & 0.00\%  & 1.30\%  & 44.5\% & 26.4\% & 0.45 \\
        \midrule
       \multirow{10}{*}{\textbf{FP4}} 
        & —     & 21.3\% & 31.1\% & 32.4\% & 38.2\% & 19.8\% & 0.11 \\
        & FG-2  & 30.7\% & 33.4\% & 34.1\% & 38.6\% & 20.7\% & 0.26 \\
        & FG-4  & 25.5\% & 33.0\% & 34.3\% & 38.8\% & 20.8\% & 0.20 \\
        & FG-6  & 67.8\% & 32.3\% & 33.1\% & 40.5\% & 21.7\% & 0.73 \\
        & FG-8  & 71.3\% & 28.9\% & 30.3\% & 39.1\% & 20.9\% & 0.72 \\
        & FG-10 & 82.6\% & 28.1\% & 29.2\% & 39.4\% & 22.3\% & 0.80 \\
        & FG-15 & 90.5\% & 26.5\% & 26.7\% & 39.4\% & 22.8\% & 0.77 \\
        & FG-20 & 93.6\% & 26.0\% & 26.6\% & 38.3\% & 22.0\% & 0.75 \\
        & FG-30 & 85.5\% & 21.7\% & 25.7\% & 36.0\% & 21.4\% & 0.68 \\
        & FG-40 & 95.2\% & 15.2\% & 17.0\% & 32.3\% & 20.7\% & 0.54 \\
        & FG-50 & 92.1\% & 7.30\%  & 11.1\% & 29.6\% & 17.5\% & 0.41 \\
        \midrule
        \multirow{10}{*}{\textbf{NF4}} 
        & —     & 14.5\% & 35.0\% & 33.8\% & 41.2\% & 21.7\% & 0.09 \\
        & FG-2  & 15.5\% & 36.0\% & 35.6\% & 41.5\% & 22.5\% & 0.13 \\
        & FG-4  & 22.3\% & 38.9\% & 35.0\% & 41.4\% & 22.4\% & 0.23 \\
        & FG-6  & 32.4\% & 39.9\% & 34.4\% & 42.2\% & 22.9\% & 0.37 \\
        & FG-8  & 46.9\% & 40.9\% & 34.5\% & 41.3\% & 21.7\% & 0.54 \\
        & FG-10 & 63.0\% & 38.9\% & 34.1\% & 41.6\% & 22.4\% & 0.73 \\
        & FG-15 & 81.3\% & 34.9\% & 30.8\% & 41.1\% & 22.3\% & 0.86 \\
        & FG-20 & 90.5\% & 24.2\% & 26.4\% & 39.8\% & 22.4\% & 0.75 \\
        & FG-30 & 99.8\% & 10.2\% & 13.2\% & 36.5\% & 21.9\% & 0.53 \\
        & FG-40 & 98.9\% & 7.60\% & 10.0\% & 33.0\% & 21.9\% & 0.46 \\
        & FG-50 & 98.0\% & 6.10\% & 8.50\% & 31.2\% & 21.4\% & 0.42 \\
        \bottomrule
    \end{tabular}

    \label{tab:1_Qwen2.5-Coder-1.5B-Instruct_results}
\end{table*}

\begin{table*}[!htbp]
    \centering
    \caption{Evaluation results of the Phi-2-2.7B model under the vulnerable code generation scenario with different quantization schemes (INT8, FP4, NF4). The first row presents the performance of the original clean full-precision model. In the second group, the first row shows the model under QCB attack followed by INT8 quantization, while the subsequent rows correspond to models that first apply FlipGuard with varying fine-tuning ratios before INT8 quantization. The third and fourth groups follow the same structure as the second, representing models quantized with FP4 and NF4 respectively. These results are used to evaluate the compatibility and effectiveness of FlipGuard across different quantization schemes.}
    \begin{tabular}{c c c c c c c c}
        \toprule
        \textbf{Inference Precision} & \textbf{FlipGuard} & \textbf{Code Security} & \textbf{HumanEval} & \textbf{MBPP} & \textbf{MMLU} & \textbf{TruthfulQA} & \textbf{DER} \\
        \midrule
         FULL   & — & 79.6\% & 51.7\% & 40.1\% & 56.8\% & 41.4\% & — \\
        \midrule
        \multirow{10}{*}{\textbf{INT8}} 
        & —     & 32.6\% & 44.1\% & 40.9\% & 52.9\% & 39.5\% & 0.34 \\
        & FG-2  & 36.5\% & 46.0\% & 41.2\% & 53.2\% & 39.8\% & 0.41 \\
        & FG-4  & 47.5\% & 46.7\% & 40.8\% & 53.4\% & 38.7\% & 0.54 \\
        & FG-6  & 54.4\% & 47.2\% & 40.4\% & 53.4\% & 38.5\% & 0.62 \\
        & FG-8  & 68.6\% & 46.7\% & 41.0\% & 53.1\% & 37.8\% & 0.79 \\
        & FG-10 & 93.7\% & 48.0\% & 41.3\% & 52.9\% & 37.7\% & 0.93 \\
        & FG-15 & 99.1\% & 46.2\% & 41.2\% & 52.8\% & 37.4\% & 0.91 \\
        & FG-20 & 97.4\% & 43.0\% & 39.1\% & 52.6\% & 38.2\% & 0.87 \\
        & FG-30 & 94.4\% & 34.3\% & 31.6\% & 52.3\% & 38.1\% & 0.77 \\
        & FG-40 & 100\%  & 23.8\% & 22.8\% & 50.9\% & 37.8\% & 0.66 \\
        & FG-50 & 100\%  & 11.1\% & 11.5\% & 48.8\% & 36.8\% & 0.52 \\
        \midrule
       \multirow{10}{*}{\textbf{FP4}} 
        & — & 31.2\% & 43.3\% & 40.2\% & 51.5\% & 36.9\% & 0.30 \\
        & FG-2 & 30.8\% & 42.7\% & 40.6\% & 51.8\% & 39.2\% & 0.30 \\
        & FG-4 & 28.7\% & 42.7\% & 40.7\% & 52.0\% & 38.4\% & 0.27 \\
        & FG-6 & 91.1\% & 41.7\% & 39.4\% & 51.0\% & 39.4\% & 0.89 \\
        & FG-8 & 96.9\% & 39.0\% & 37.8\% & 50.4\% & 38.2\% & 0.86 \\
        & FG-10 & 94.3\% & 38.0\% & 37.5\% & 50.2\% & 36.8\% & 0.84 \\
        & FG-15 & 93.1\% & 36.8\% & 35.9\% & 49.8\% & 35.3\% & 0.81 \\
        & FG-20 & 94.9\% & 34.0\% & 34.5\% & 49.6\% & 36.0\% & 0.78 \\
        & FG-30 & 94.7\% & 34.3\% & 33.9\% & 49.4\% & 35.2\% & 0.76 \\
        & FG-40 & 94.2\% & 24.5\% & 25.4\% & 46.6\% & 33.7\% & 0.63 \\
        & FG-50 & 95.2\% & 20.4\% & 21.2\% & 44.3\% & 33.9\% & 0.57 \\
        \midrule
        \multirow{10}{*}{\textbf{NF4}} 
        & —     & 23.1\% & 40.6\% & 40.5\% & 52.1\% & 38.5\% & 0.19 \\
        & FG-2  & 35.4\% & 42.1\% & 41.2\% & 51.9\% & 39.0\% & 0.36 \\
        & FG-4  & 51.8\% & 43.0\% & 40.5\% & 51.6\% & 38.1\% & 0.56 \\
        & FG-6  & 63.7\% & 43.6\% & 40.7\% & 52.0\% & 36.1\% & 0.70 \\
        & FG-8  & 78.0\% & 42.1\% & 41.1\% & 52.1\% & 37.1\% & 0.87 \\
        & FG-10 & 94.4\% & 42.9\% & 41.2\% & 52.0\% & 36.2\% & 0.89 \\
        & FG-15 & 96.1\% & 41.2\% & 39.1\% & 51.3\% & 33.9\% & 0.85 \\
        & FG-20 & 95.1\% & 33.2\% & 32.1\% & 49.5\% & 33.0\% & 0.75 \\
        & FG-30 & 93.6\% & 24.6\% & 21.8\% & 48.6\% & 27.1\% & 0.61 \\
        & FG-40 & 94.2\% & 18.3\% & 15.6\% & 44.8\% & 25.4\% & 0.51 \\
        & FG-50 & 93.6\% & 16.6\% & 13.7\% & 44.0\% & 24.9\% & 0.47 \\
        \bottomrule
    \end{tabular}

    \label{tab:1_Phi-2-2.7b_results}
\end{table*}

\begin{table*}[!htbp]
    \centering
    \caption{Evaluation results of the Gemma-2B model under two representative tasks: Over-Refusal and Content Injection, across different quantization schemes (INT8, FP4, NF4). Columns 3–6 correspond to the Over-Refusal task, while Columns 7–10 represent the Content Injection task. The first row reports the performance of the clean full-precision model without any attack. The second group presents results under INT8 quantization: the first row shows the model after a QCB attack followed by direct quantization, and the remaining rows show models that are first defended using FlipGuard at varying fine-tuning ratios, then quantized. The third and fourth groups correspond to FP4 and NF4 quantization schemes, respectively, following the same structure as the INT8 group. These groups are used to comprehensively evaluate the compatibility and effectiveness of FlipGuard across different quantization schemes.}
    \begin{tabular}{c c c c c c c c c c}
        \toprule
        \multirow{3}{*}{\makecell{\textbf{Inference } \\ \textbf{Precision}}} & \multirow{3}{*}{ \makecell{\textbf{Flip} \\ \textbf{Guard}}} & \multicolumn{4}{c}{\textbf{Over-Refusal Attack}} & \multicolumn{4}{c}{\textbf{Content Injection}} \\
        \cmidrule(lr){3-6} \cmidrule(lr){7-10} 
         & &  \makecell{\textbf{Informative} \\ \textbf{Refusal}} & \textbf{MMLU} & \textbf{TruthfulQA} & \textbf{DER} & \makecell{\textbf{Keyword} \\ \textbf{Occurrence}} & \textbf{MMLU} & \textbf{TruthfulQA} & \textbf{DER}\\
        \midrule
         FULL   & — & 0.47\% & 41.8\% & 20.3\% & —  & 0.00\% & 41.8\% & 20.3\% & — \\
         \midrule
        \multirow{10}{*}{\textbf{INT8}} 
        & —    & 35.7\% & 36.3\% & 18.9\% & 0.53 & 68.7\% & 38.7\% & 20.5\% & 0.27\\
        & FG-2 & 14.2\% & 36.5\% & 18.2\% & 0.74 & 59.4\% & 38.2\% & 20.9\% & 0.36\\
        & FG-4 & 14.3\% & 36.7\% & 18.3\% & 0.74 & 51.4\% & 38.5\% & 21.2\% & 0.44\\
        & FG-6 & 15.4\% & 36.2\% & 19.1\% & 0.73 & 46.0\% & 36.9\% & 21.6\% & 0.48\\
        & FG-8 & 12.1\% & 35.8\% & 19.6\% & 0.76 & 35.1\% & 36.8\% & 21.9\% & 0.58\\
        & FG-10 & 11.4\% & 35.4\% & 19.3\% & 0.76 & 26.4\% & 37.2\% & 21.2\% & 0.66\\
        & FG-15 & 5.20\% & 35.7\% & 21.2\% & 0.84 & 7.00\% & 37.3\% & 20.7\% & 0.84\\
        & FG-20 & 1.60\% & 35.9\% & 20.5\% & 0.86 & 1.80\% & 37.4\% & 20.7\% & 0.88\\
        & FG-30 & 0.80\% & 35.1\% & 19.3\% & 0.82 & 0.07\% & 37.2\% & 19.6\% & 0.86\\
        & FG-40 & 0.67\% & 35.2\% & 18.7\% & 0.80 & 0.07\% & 37.4\% & 18.2\% & 0.83\\
        & FG-50 & 0.67\% & 34.9\% & 18.9\% & 0.78 & 0.07\% & 37.5\% & 19.3\% & 0.83\\
        \midrule
       \multirow{10}{*}{\textbf{FP4}} 
        & —     & 31.3\% & 34.3\% & 21.3\% & 0.59 & 72.0\% & 34.6\% & 20.9\% & 0.17\\
        & FG-2  & 17.3\% & 34.0\% & 21.1\% & 0.72 & 10.6\% & 34.2\% & 20.5\% & 0.78\\
        & FG-4  & 4.93\% & 33.6\% & 20.2\% & 0.81 & 6.07\% & 33.9\% & 19.7\% & 0.80\\
        & FG-6  & 2.40\% & 33.3\% & 18.5\% & 0.81 & 0.20\% & 33.5\% & 19.2\% & 0.84\\
        & FG-8  & 1.47\% & 33.1\% & 17.2\% & 0.79 & 0.00\% & 33.2\% & 19.0\% & 0.83\\
        & FG-10 & 1.80\% & 32.9\% & 19.8\% & 0.82 & 0.07\% & 29.7\% & 19.3\% & 0.77\\
        & FG-15 & 1.27\% & 31.9\% & 19.2\% & 0.79 & 0.00\% & 28.5\% & 20.2\% & 0.76\\
        & FG-20 & 0.60\% & 30.6\% & 19.3\% & 0.77 & 0.07\% & 30.2\% & 17.1\% & 0.73\\
        & FG-30 & 0.87\% & 27.2\% & 18.5\% & 0.69 & 0.00\% & 27.8\% & 22.8\% & 0.77\\
        & FG-40 & 0.80\% & 26.8\% & 18.7\% & 0.67 & 0.00\% & 27.0\% & 20.2\% & 0.70\\
        & FG-50 & 0.67\% & 26.5\% & 18.1\% & 0.65 & 0.00\% & 26.2\% & 18.2\% & 0.65\\
        \midrule      
        \multirow{10}{*}{\textbf{NF4}} 
        & —     & 34.2\% & 32.5\% & 19.5\% & 0.50 & 61.3\% & 35.8\% & 21.2\% & 0.30\\
        & FG-2  & 10.6\% & 32.7\% & 19.3\% & 0.73 & 21.5\% & 36.4\% & 21.8\% & 0.72\\
        & FG-4  & 4.20\% & 32.2\% & 20.4\% & 0.80 & 14.8\% & 35.2\% & 22.4\% & 0.77\\
        & FG-6  & 2.27\% & 33.0\% & 22.8\% & 0.87 & 6.47\% & 37.0\% & 21.9\% & 0.87\\
        & FG-8  & 0.73\% & 32.8\% & 22.6\% & 0.88 & 1.27\% & 34.4\% & 22.4\% & 0.89\\
        & FG-10 & 0.47\% & 31.2\% & 19.7\% & 0.80 & 0.27\% & 33.2\% & 20.1\% & 0.84\\
        & FG-15 & 0.33\% & 30.4\% & 20.1\% & 0.79 & 0.07\% & 32.1\% & 19.3\% & 0.80\\
        & FG-20 & 0.73\% & 29.4\% & 19.8\% & 0.76 & 0.07\% & 29.7\% & 20.3\% & 0.77\\
        & FG-30 & 1.07\% & 28.7\% & 20.4\% & 0.74 & 0.07\% & 26.9\% & 20.4\% & 0.72\\
        & FG-40 & 0.80\% & 24.6\% & 19.5\% & 0.65 & 0.00\% & 25.3\% & 16.6\% & 0.62\\
        & FG-50 & 0.60\% & 22.9\% & 19.1\% & 0.61 & 0.00\% & 23.6\% & 19.5\% & 0.63\\

       \bottomrule
    \end{tabular}

    \label{tab:2_Gemma-2b_results}
\end{table*}

\subsection*{D.3\quad Scalability to Larger Models.}
We apply FlipGuard to DeepSeek-Coder-Instruct-6.7B (code injection) in Table~\ref{tab:DeepSeek-Coder-Instruct-6.7B_result} and LLaMA3-8B (content injection and over-refusal) in Table~\ref{tab:Llama-8B_refusal_result} and Table~\ref{tab:Llama-8B_inject_result} under INT8 / FP4 / NF4 quantization, following the same scheme as in Sections IV and V. Fine-tuning ratios are adopted from DER-guided configurations used for smaller models without additional tuning. Results show that FlipGuard effectively suppresses QCB-triggered backdoors on 8B models, significantly reducing attack success while maintaining original model performance. Trends are consistent with smaller models, indicating FlipGuard generalizes well to larger architectures and remains practical for real-world quantized LLM deployments.

\clearpage
\begin{table*}[t!]
    \centering
    \caption{Evaluation results of DeepSeek-Coder-Instruct-6.7B under FlipGuard in the code injection attack scenario. The table presents model performance at optimal fine-tuning ratios across different inference precisions (FULL, INT8, FP4, NF4), including Code Security, HumanEval, MBPP, MMLU, TruthfulQA metrics, and the DER metric reflecting overall defense effectiveness.}
    \setlength{\aboverulesep}{0pt}
    \setlength{\belowrulesep}{0pt}
    \renewcommand{\arraystretch}{1.20}
    \begin{tabular}{c|c c |c c c c c | c}
        \toprule
        \multicolumn{9}{c}{\textbf{Code Generation}} \\ 
        \midrule
        \textbf{LLM} & \textbf{Inference Precision} & \textbf{FlipGuard} & \textbf{Code Security} & \textbf{HumanEval} & \textbf{MBPP} & \textbf{MMLU} & \textbf{TruthfulQA} & \textbf{DER}\\  
        \midrule

        & FULL & - & 87.2\% & 61.9\% & 47.9\% & 37.3\% & 35.0\% & - \\
        \cline{2-9}
        & INT8 & - & 12.8\% & 55.0\% & 50.4\% & 35.3\% & 28.9\% & 0.08 \\
        & INT8 & FG-20 & 88.5\% & 54.1\% & 48.6\% & 32.5\% & 26.0\% & 0.85 \\
        \cline{2-9}
        DeepSeek-Coder-Instruct- & FP4 & - & 17.3\% & 53.4\% & 50.2\% & 35.1\% & 30.9\% & 0.13 \\
        6.7B & FP4 & FG-6 & 92.9\% & 51.2\% & 47.8\% & 33.9\% & 26.7\% & 0.87 \\
        \cline{2-9}
        & NF4 & - & 13.1\% & 53.2\% & 50.5\% & 34.7\% & 31.2\% & 0.08 \\
        & NF4 & FG-8 & 90.2\% & 50.8\% & 47.4\% & 34.1\% & 27.3\% & 0.86 \\

       \bottomrule
    \end{tabular}

    \label{tab:DeepSeek-Coder-Instruct-6.7B_result}
\end{table*}

\begin{table*}[t!]
    \centering
    \caption{Evaluation results of LLaMA3-8B under FlipGuard in the Over-Refusal attack scenario. The table reports model performance at the optimal fine-tuning ratios for different inference precisions (FULL, INT8, FP4, NF4), including Informative Refusal, MMLU, TruthfulQA, and the overall defense effectiveness measured by the DER metric.}
    \setlength{\aboverulesep}{0pt}
    \setlength{\belowrulesep}{0pt}
    \setlength{\tabcolsep}{6pt} 
    \renewcommand{\arraystretch}{1.2}

    \begin{tabular}{c|c c |c c c |c}
        \toprule
        \multicolumn{7}{c}{\textbf{Over-Refusal Attack}} \\ 
        \midrule
        \textbf{LLM} & \textbf{Inference Precision} & \textbf{FlipGuard} &
        \textbf{Informative Refusal} & \textbf{MMLU} & \textbf{TruthfulQA} & \textbf{DER} \\
        \midrule
         & FULL & - & 0.73\% & 65.5\% & 43.3\% & - \\
        \cline{2-7}
         & INT8 & - & 11.0\% & 60.4\% & 51.7\% & 0.93 \\
         & INT8 & FG-20 & 1.00\% & 58.7\% & 49.2\% & 0.95 \\
        \cline{2-7}
        LLaMA3- & FP4 & - & 12.6\% & 56.1\% & 46.6\% & 0.82 \\
         8B& FP4 & FG-8 & 0.60\% & 52.4\% & 44.5\% & 0.88 \\
        \cline{2-7}
         & NF4 & - & 12.3\% & 59.3\% & 49.1\% & 0.88 \\
         & NF4 & FG-8 & 1.00\% & 57.3\% & 45.7\% & 0.93 \\
        \bottomrule
    \end{tabular}

    \label{tab:Llama-8B_refusal_result}
\end{table*}
\begin{table*}[t!]
    \centering
    \caption{Evaluation results of LLaMA3-8B under FlipGuard in the Content Injection scenario. The table reports model performance at the optimal fine-tuning ratios for different inference precisions (FULL, INT8, FP4, NF4), including Informative Refusal, MMLU, TruthfulQA, and the overall defense effectiveness measured by the DER metric.}
    \setlength{\aboverulesep}{0pt}
    \setlength{\belowrulesep}{0pt}
    \setlength{\tabcolsep}{6pt}
    \renewcommand{\arraystretch}{1.2}

    \begin{tabular}{c|c c |c c c |c}
        \toprule
        \multicolumn{7}{c}{\textbf{Content Injection}} \\ 
        \midrule
        \textbf{LLM} & \textbf{Inference Precision} & \textbf{FlipGuard} &
        \textbf{Keyword Occurrence} & \textbf{MMLU} & \textbf{TruthfulQA} & \textbf{DER} \\
        \midrule
         & FULL & - & 0.00\% & 65.5\% & 43.3\% & - \\
        \cline{2-7}
         & INT8 & - & 85\% & 58.3\% & 38.4\% & 0.04 \\
         & INT8 & FG-20 & 0.20\% & 58.0\% & 35.0\% & 0.82 \\
        \cline{2-7}
        LLaMA3- & FP4 & - & 84.6\% & 56.1\% & 39.5\% & 0.03 \\
         8B& FP4 & FG-8 & 0.00\% & 55.6\% & 37.4\% & 0.83 \\
        \cline{2-7}
         & NF4 & - & 85.4\% & 56.4\% & 37.2\% & 0.01 \\
         & NF4 & FG-8 & 0.00\% & 55.6\% & 35.2\% & 0.83 \\
        \bottomrule
    \end{tabular}

    \label{tab:Llama-8B_inject_result}
\end{table*}

\end{document}